\RequirePackage{pdf14}

\pdfoutput=1
\documentclass[letterpaper]{article}
\usepackage[utf8]{inputenc}
\usepackage[usenames,dvipsnames,table]{xcolor}
\usepackage{graphicx}
\usepackage{caption}
\usepackage{subcaption}
\usepackage{array,setspace,mathrsfs,amsthm,amsfonts,amsmath,colonequals,mathtools,mathrsfs,xspace,pgfplots,dsfont,tikz}
\usepackage{./auxi/jheppub}
\graphicspath{{./figures/}}

\pgfplotsset{compat=1.7}

\let\oldbfseries=\bfseries
\let\oldmdseries=\mdseries
\let\oldnormalfont=\normalfont
\renewcommand{\bfseries}{\oldbfseries\boldmath}
\renewcommand{\mdseries}{\oldmdseries\unboldmath}
\renewcommand{\normalfont}{\oldnormalfont\unboldmath}

\makeatletter
\newlength{\apb@width}
\newcommand{\autoparbox}[2][c]{\settowidth{\apb@width}{#2}\parbox[#1]{\apb@width}{#2}}

\makeatother

\newcommand{\so}{\ensuremath{\mathfrak{so}}}%
\newcommand{\su}{\ensuremath{\mathfrak{su}}}%
\newcommand{\osp}{\ensuremath{\mathfrak{osp}}}%
\newcommand{\usp}{\ensuremath{\mathfrak{usp}}}%
\renewcommand{\sl}{\ensuremath{\mathfrak{sl}}}%

\newcommand{\BB}{\ensuremath{\mathcal{B}}}%
\newcommand{\CC}{\ensuremath{\mathcal{C}}}%
\newcommand{\DD}{\ensuremath{\mathcal{D}}}%
\newcommand{\GG}{\ensuremath{\mathcal{G}}}%
\newcommand{\II}{\ensuremath{\mathcal{I}}}%
\newcommand{\PP}{\ensuremath{\mathcal{P}}}%
\newcommand{\KK}{\ensuremath{\mathcal{K}}}%
\newcommand{\LL}{\ensuremath{\mathcal{L}}}%
\newcommand{\MM}{\ensuremath{\mathcal{M}}}%
\newcommand{\NN}{\ensuremath{\mathcal{N}}}%
\newcommand{\OO}{\ensuremath{\mathcal{O}}}%
\newcommand{\RR}{\ensuremath{\mathcal{R}}}%
\newcommand{\QQ}{\ensuremath{\mathcal{Q}}}%
\renewcommand{\SS}{\ensuremath{\mathcal{S}}}%
\newcommand{\VV}{\ensuremath{\mathcal{V}}}%

\newcommand{\bA}{\ensuremath{\mathbf{A}}}%
\newcommand{\bB}{\ensuremath{\mathbf{B}}}%
\newcommand{\bC}{\ensuremath{\mathbf{C}}}%
\newcommand{\bD}{\ensuremath{\mathbf{D}}}%
\newcommand{\bs}[1]{\ensuremath{\boldsymbol{#1}}}%
\newcommand{\ia}{\ensuremath{a}}%
\newcommand{\ib}{\ensuremath{b}}%
\newcommand{\ic}{\ensuremath{c}}%
\newcommand{\id}{\ensuremath{d}}%
\newcommand{\dispM}{\DD}%
\newcommand{\longM}{\OO}%

\def \ph{\phantom}

\newcommand{\ee}{\ensuremath{\varepsilon}} % Epsilon symbol
\newcommand{\dd}{\ensuremath{\mathrm{d}}} % Differential
\newcommand{\px}{\ensuremath{\partial_x}} % Partial derivative w.r.t. x
\newcommand{\pt}[2]{\ensuremath{\partial^{#1}_{#2}}} % Partial derivative w.r.t. theta
\newcommand{\cc}{\ensuremath{\mathfrak c}} % Casimir eigenvalue
\newcommand{\fr}[1]{\ensuremath{\mathscr{F}_{#1}}} % Frames
\newcommand{\Rsymmetry}{$R$-symmetry\xspace}

\def\Om{{\mathcal{O}}}

\def\Nm{{\mathcal{N}}}

\def\Dm{{\mathcal{D}}}

\def\veps{\varepsilon}
\def\ph{\phantom}
\def\pd{\partial}

\def\veps{\varepsilon}
\def\ph{\phantom}
\def\pd{\partial}

\newcommand{\beq}{\begin{equation}}
\newcommand{\eeq}{\end{equation}}

\newcommand{\mathematica}{\texttt{Mathematica}\xspace}

\newmuskip\pFqskip
\mathchardef\pFcomma=\mathcode`,

\title{Bootstrapping line defects in $\Nm=2$ theories}
\author{Aleix Gimenez-Grau,}
\author{Pedro Liendo.}

\affiliation{DESY Hamburg, Theory Group, Notkestra{\ss}e 85, D-22607 Hamburg, Germany}

\emailAdd{aleix.gimenez@desy.de}
\emailAdd{pedro.liendo@desy.de}

\preprint{DESY 19-126}

\bigskip
\abstract{
We study half-BPS line defects in $\Nm=2$ superconformal theories using the bootstrap approach. We concentrate on local excitations constrained to the defect, which means the system is a $1d$ defect CFT with $\osp(4^*|2)$ symmetry.
In order to study correlation functions we construct a suitable superspace, and then use the Casimir approach to calculate a collection of new superconformal blocks. Special emphasis is given to the displacement operator, which controls deformations orthogonal to the defect and is always present in a defect CFT. 
After setting up the crossing equations we proceed with a numerical and analytical bootstrap analysis. We obtain numerical bounds on the CFT data and compare them to known solutions. We also present an analytic perturbative solution to the crossing equations, and argue that this solution captures line defects in $\Nm=2$ gauge theories at strong coupling.
}


\begin{document}
\setcounter{tocdepth}{2}
\maketitle
\setcounter{page}{1}

%!TEX root = ../N2_line.tex
%%%%%%%%%%%%%%%%%%%%%%%%%%%%%%%%%%%%%%%%%%%%%
\section{Introduction}

Defects are important observables in quantum field theory: they serve as probes that allow to extract physics otherwise inaccessible from the study of local operators.
In four-dimensional gauge theories, it is well understood by now that models with the same local correlators might have different line operators, and therefore correspond to distinct physical theories \cite{Aharony:2013hda}. In this work, we concentrate on line defects in $4d$ superconformal theories with $\Nm=2$ supersymmetry. In particular, we consider half-BPS defects that preserve an $\osp(4^*|2)$ subalgebra of the full $\su(2,2|2)$ superconformal algebra.

An important example of such a defect is a Wilson line operator, which describes a charged heavy particle moving in the vacuum of a gauge theory. Due to the high amount of supersymmetry preserved by the configuration, it is possible to obtain exact formulas using localization and related matrix model techniques \cite{Pestun:2007rz}.
For example, the Bremsstrahlung function, which captures the energy radiated by the particle, can be calculated exactly \cite{Correa:2012at,Lewkowycz:2013laa,Fiol:2012sg,Fiol:2015spa}. A way to understand this is that the Bremsstrahlung is proportional to the one-point function of the stress tensor in the presence of the line, and the latter can be obtained from localization. This relation between Bremsstrahlung and the stress tensor was conjectured in \cite{Fiol:2015spa} for $\Nm=2$ theories, and later proven in \cite{Bianchi:2018zpb} using only superconformal symmetry.

The literature on Wilson operators in $\Nm=2$ theories is vast, however work on configurations with insertions along the contour has been scarce. Here we study this system from the $1d$ CFT perspective by analyzing correlators of operators inserted on the line. Although $1d$ theories are non-local due to the absence of a stress tensor, they are consistent when interpreted as defect theories. Correlators on a defect can be thought of as describing a lower dimensional CFT embedded in a higher dimensional bulk. In particular, four-point functions exhibit crossing symmetry and have a conformal block expansion with positive coefficients. Thanks to this positivity property, one can use the numerical bootstrap of \cite{Rattazzi:2008pe} to constrain the CFT data.  We should mention that if one considers operators outside the defect the positivity property is lost, and the numerical bootstrap does not apply. One can nevertheless use analytical bootstrap techniques, see \cite{Lemos:2017vnx,Liendo:2019jpu} for recent progress.

The canonical operator that is always present on a defect CFT is the displacement operator. This operator measures deformations orthogonal to the defect, and is the closest one can have to a conserved stress tensor. Indeed, the stress tensor and the displacement are related by a Ward identity \cite{Billo:2016cpy}. Due to its universal character, in this work we concentrate on the four-point function of the displacement operator. Because the system we are considering is supersymmetric, in order to study the displacement operator, it will be necessary to study the corresponding superconformal multiplet. Our bootstrap analysis is based on symmetry and we will not commit to any particular theory. This work is complementary to the bulk $\Nm=2$ superconformal bootstrap program \cite{Beem:2014zpa,Lemos:2015awa,Cornagliotto:2017snu}, where the main focus is the study of correlators of local operators.\footnote{We should also mention that $\Nm=2$ theories admit a wide variety of codimension-2 surface operators, but here we only concentrate on codimension-3 defects.}

In $\Nm=4$ SYM, the corresponding line defect with insertions has been studied recently using a variety of techniques. These include explicit holographic calculations \cite{Giombi:2017cqn}, the conformal bootstrap \cite{Liendo:2016ymz,Liendo:2018ukf}, truncations to the topological sector \cite{Giombi:2018qox,Giombi:2018hsx}, and perturbative calculations at weak coupling \cite{Cooke:2017qgm,Cooke:2018obg}.
Another related system is the monodromy line of the $3d$ Ising model \cite{Billo:2013jda}, which was studied using bootstrap techniques in \cite{Gaiotto:2013nva}.
Apart from their intrinsic interest, $1d$ CFTs are also a useful laboratory in which bootstrap ideas can be explored. Recent work includes exact functionals that allow to extract the spectrum analytically \cite{Mazac:2016qev,Mazac:2018mdx,Mazac:2018ycv}, inversion formulas \cite{Simmons-Duffin:2017nub,Mazac:2018qmi} (see also \cite{Bissi:2018mcq,Kaviraj:2018tfd}  for the closely related case of BCFT), and intriguing positivity properties \cite{Arkani-Hamed:2018ign}. 

The structure of the paper is as follows. In section~\ref{sec:preliminaries} we review the geometry of our setup and present the preserved $\osp(4^*|2)$ superconformal algebra.
We find all its unitary representations and explicitly construct the multiplets of long and short operators that will play a role in later discussions.
In section~\ref{sec:superspace} we construct correlation functions using superspace, concentrating on those containing the multiplet of the displacement operator.
With the superspace at hand, in section \ref{sec:superblocks} we use the Casimir approach to calculate the superconformal blocks involving four displacement multiplets.
We write the associated crossing equations, and find a solution that interpolates between bosonic and fermionic free-field theory.
We apply standard numerical bootstrap techniques to our crossing equations in section~\ref{sec:numerical}, and we find that the free-field solutions sit in interesting points of the allowed regions of the plots, where they saturate the numerical bounds.
In section~\ref{sec:analytical} we employ analytic techniques to find a solution to crossing which we interpret as a perturbative first-order correction to the strong-coupling limit of our line defect.
Finally, we conclude in section~\ref{sec:conclusions} by giving an outlook on possible future directions of research.
We complement the text with our conventions (appendix~\ref{apx:conventions}), and a compendium of superconformal blocks of unprotected long operators (appendix~\ref{apx:long-blocks}), which can be useful in future studies of this setup.
We also attach a \mathematica file with a number of technical results.

%!TEX root = ../N2_line.tex
%%%%%%%%%%%%%%%%%%%%%%%%%%%%%%%%%%%%%%%%%%%%%

\section{Preliminaries}
\label{sec:preliminaries}

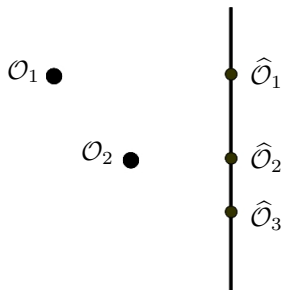
\begin{figure}[ht!]
	\centering
	\includegraphics[scale=0.3]{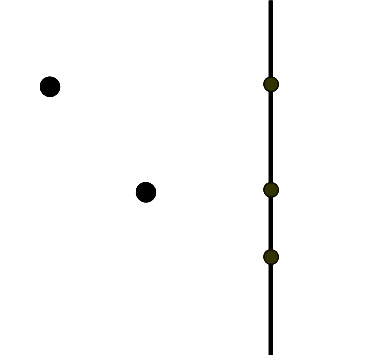}
	\put(-115,80){$\Om_1$}
	\put(-87,49){$\Om_2$}
	\put(-23,78){$\widehat{\Om}_1$}
	\put(-23,46){$\widehat{\Om}_2$}
	\put(-23,24){$\widehat{\Om}_3$}
	\caption{In the presence of a defect, one can consider correlators of local and defect operators. Because the defect breaks the conformal algebra down to a subalgebra, even low-point functions can acquire non-trivial coordinate dependence. In this work we will concentrate exclusively on defect excitations (hatted operators in the figure) which define a lower dimensional CFT.}
	\label{fig:line_defect}
\end{figure}

There are several configurations one can consider when studying defect CFTs: correlation functions of local operators in the presence of the defect, correlators of defect operators, i.e. local excitations that are constrained to live on the defect, and also mixed configurations with both local and defect operators (see figure \ref{fig:line_defect}). 
Because defects break some of the conformal symmetry, even low-point correlators tend to have non-trivial structure. One-point functions of local operators are generically non-zero, and two-point functions have a non-trivial dependence on two conformal invariants \cite{Billo:2016cpy}, which makes them analogous to four-point functions in bulk CFTs with no defects. 

In this work we will study line defects in four dimensions, and we concentrate exclusively on defect excitations. 
 We will consider correlators of the canonical operator that is always present on a defect CFT: the \textit{displacement operator}. This universal operator measures deformations orthogonal to the defect. Intuitively, it can be thought of as the orthogonal components of the stress tensor, which is the generator of translations. 
Since we are resticting ourselves to the line, our system is described by a $1d$ CFT and all the usual bootstrap techniques apply\footnote{See appendix A of~\cite{Qiao:2017xif} for a general introduction to $1d$ CFTs.}. In particular, four-point functions have a conformal block expansion with positive coefficients and they satisfy a crossing symmetry equation.
 
The symmetry algebra preserved by our defect is $\osp(4^*|2)$\footnote{A very complete presentation of superalgebras and their real forms can be found in~\cite{Frappat:1996pb}.}, which is a subalgebra of the full $\Nm=2$ superconformal algebra. This is the maximal possible superalgebra consistent with the geometry of the configuration. In Lagrangian theories, special boundary conditions can be chosen in order to preserve $\osp(4^*|2)$, but here we will not consider any particular model and we rely only on algebraic and symmetry constraints: the $\osp(4^*|2)$ symmetry algebra will be our starting point.

In four dimensions a line defect has three orthogonal directions, and therefore the displacement is a vector. In the supersymmetric setup we are considering, the displacement sits in a supermultiplet whose highest weight is a scalar. This means that after taking into account all the constraints coming from supersymmetry, our analysis will be similar to the $1d$ bosonic bootstrap. In the next subsection we review the $\osp(4^*|2)$ superalgebra together with its representation theory, with special emphasis on the multiplets which will be relevant when studying crossing symmetry in section \ref{sec:superblocks}.

\subsection{The superalgebra}

We are interested in line defects that preserve the maximum amount of supersymmetry $\osp(4^*|2)$, with bosonic subalgebra $\sl(2; \mathbb R) \oplus \su(2)_j \oplus \su(2)_R$.  
In addition to the $\sl(2; \mathbb R)$ factor which captures the $1d$ conformal symmetry, there is an extra $\so(3) \cong \su(2)_j$ which can be interpreted as rotations around the defect. 
The quantum number associated to it, which we label by $j$, is called transverse spin.
The last $\usp(2) \cong \su(2)_R$ is the leftover \Rsymmetry preserved by the configuration. 
For transverse-spin indices we will use $\ia = 1,2$, and for \Rsymmetry indices $\bA = 1,2$. 
The fermionic generators are given by supercharges $\QQ$ and $\SS$, and carry both types of indices. 
The bosonic part of the superalgebra is given by
\begin{align}
\label{eq:osp42_superalgebra_bos}
\begin{split}
[\DD, \PP] & =   \PP, \\
[\DD, \KK] & = - \KK, \\
[\KK, \PP] & = 2 \DD, \\
[\MM_\ia^{\ph \ia\ib}, \MM_\ic^{\ph \ic\id}] 
& = - \delta_\ia^{\ph \ia\id} \MM_\ic^{\ph \ic\ib}
+ \delta_\ic^{\ph \ic\ib} \MM_\ia^{\ph \ia\id}, \\
[\RR^\bA_{\ph \bA\bB}, \RR^\bC_{\ph \bC\bD}]
& = - \delta^\bA_{\ph \bA\bD} \RR^\bC_{\ph \bC\bB}
+ \delta^\bC_{\ph \bC\bB} \RR^\bA_{\ph \bA\bD}.
\end{split}
\end{align}
The fermionic generators anticommute as follows
\begin{align}
\label{eq:osp42_superalgebra_ferm}
\begin{split}
\{ \QQ^\bA_\ia, \QQ^\bB_\ib \} & = \ee^{\bA \bB} \ee_{\ia\ib} \PP, \\
\{ \SS_\bA^\ia, \SS_\bB^\ib \} & = \ee_{\bA \bB} \ee^{\ia\ib} \KK, \\
\{ \QQ^\bA_\ia, \SS_\bB^\ib \} 
& = - 2 \delta_\ia^{\ph \ia\ib} \RR^\bA_{\ph \bA\bB} 
+ \delta^\bA_{\ph \bA\bB} ( \MM_\ia^{\ph \ia\ib} + \delta_\ia^{\ph \ia\ib} \DD).
\end{split}
\end{align}
Finally, the fermionic generators have the following commutation relations with the bosonic subalgebra
\begin{align}
\label{eq:osp42_superalgebra_bosferm}
\begin{split}
[\DD, \QQ^\bA_\ia] & = \tfrac{1}{2} \QQ^\bA_\ia, \\
[\PP, \QQ^\bA_\ia] & = 0, \\
[\KK, \QQ^\bA_\ia] & = \ee^{\bA \bB} \ee_{\ia\ib} \SS^\ib_\bB, \\
[\MM_\ia^{\ph \ia\ib}, \QQ^\bC_\ic] 
& = \delta_\ic^{\ph \ic\ib} \QQ^\bC_\ia - \tfrac{1}{2} \delta_\ia^{\ph \ia\ib} \QQ^\bC_\ic, \\
[\RR^\bA_{\ph \bA\bB}, \QQ^\bC_\ic] 
& = \delta^\bC_{\ph \bC\bB} \QQ^\bA_\ic - \tfrac{1}{2} \delta^\bA_{\ph \bA\bB} \QQ^\bC_\ic,
\end{split}
\begin{split}
[\DD, \SS_\bA^\ia] & = - \tfrac{1}{2} \SS_\bA^\ia, \\
[\PP, \SS_\bA^\ia] & = - \ee_{\bA \bB} \ee^{\ia\ib} \QQ^\bB_\ib, \\
[\KK, \SS_\bA^\ia] & = 0, \\
[\MM_\ia^{\ph \ia\ib}, \SS_\bC^\ic] 
& = -\delta_\ia^{\ph \ia\ic} \SS_\bC^\ib + \tfrac{1}{2} \delta_\ia^{\ph \ia\ib} \SS_\bC^\ic, \\
[\RR^\bA_{\ph \bA\bB}, \SS_\bC^\ic] 
& = -\delta^\bA_{\ph \bA\bC} \SS_\bB^\ic + \tfrac{1}{2} \delta^\bA_{\ph \bA\bB} \SS_\bC^\ic.
\end{split}
\end{align}
The above superalgebra is compatible with the natural hermitian conjugation in radial quantization
\begin{align}
\begin{split}
\DD^\dag = \DD, \quad
\PP^\dag = \KK, \quad
\left(\MM_\ia^{\ph \ia\ib}\right)^\dag = \MM_\ib^{\ph \ib\ia}, \quad
\left(\RR^\bA_{\ph \bA\bB}\right)^\dag = \RR^\bB_{\ph \bB\bA}, \quad
\left(\QQ_\ia^{\bA}\right)^\dag = \SS_\bA^\ia.
\end{split}
\end{align}

\subsection{Unitary multiplets}
\label{sec:shortening-conditions}

Let us now turn to the study of unitary representations of $\osp(4^*|2)$.
The multiplet that contains the displacement operator has been constructed in~\cite{Bianchi:2018zpb}, and a similar analysis for the case of Wilson loops in ABJM can be found in~\cite{Bianchi:2017ozk,Bianchi:2018scb}.
Unitary representations of $\osp(4^*|2)$ have been previously discussed in \cite{Gunaydin:1990ag}, although here we give a more complete treatment following the work of \cite{Cordova:2016emh}.

Highest-weight representations of superconformal algebras are constructed starting from a superconformal primary field $\VV$, which is anhilated by the $\KK$ and $\SS$ generators, and transforms in some representation of the bosonic subalgebra.
For the case of interest to us, we label the primary by $[\Delta, j, R]$, where $\Delta$ is the conformal dimension, and $j,R$  are positive half-integers that label the transverse spin and \Rsymmetry respectively.
Acting with $\QQ$ supercharges on $\VV$, one obtains the conformal descendants, which are conformal primary fields, i.e. fields anhilated by the $\KK$ generator.
It is then clear that the conformal descendants form representations of the conformal algebra (but not of the superconformal algebra) on their own.
Requiring positivity of the the norm of these descendants at levels 1 and 2, imposes the unitarity bounds and shortening conditions summarized in table~\ref{tab:n2-shortening}.
To our knowledge, these results have not been presented systematically elsewhere, but we do not derive them here.
Instead, we refer the reader to the works~\cite{Minwalla:1997ka,Cordova:2016emh}, which give a detailed treatment on how to obtain unitarity bounds for all superconformal theories in $d \ge 3$.

{
	\renewcommand{\arraystretch}{1.5}
	\renewcommand\tabcolsep{8pt}
	\begin{table}
		\centering
		\begin{tabular}{ | c | l | l | l | }
			\hline
			{\bf Name} & 
			{\bf Primary} & 
			{\bf Unitarity Bound} & 
			{\bf Null State} \\ \hline \hline
			% Long
			$L$ & 
			$[\Delta, j, R]$ & 
			$\Delta > 2R + j + 1$ &
			\multicolumn{1}{c|}{$-$} \\ \hline \hline
			% Short A1
			$A_1$ & 
			$[\Delta, j, R], \, j > 0$ & 
			$\Delta = 2R + j + 1$ &
			$[\Delta + \frac{1}{2}, j - \frac{1}{2}, R + \frac{1}{2}]$ \\ \hline 
			% Short A2
			$A_2$ & 
			$[\Delta, 0, R]$ & 
			$\Delta = 2R + 1$ &
			$[\Delta + 1, 0, R+1]$  \\ \hline \hline
			% Short B1
			$B_1$ & 
			$[\Delta, 0, R]$ & 
			$\Delta = 2R$ & 
			$[\Delta + \frac{1}{2}, \frac{1}{2}, R+\frac{1}{2}]$ \\ \hline
		\end{tabular} 
		\caption{Shortening conditions in one-dimensional $\NN = 2$ SCFTs.}
		\label{tab:n2-shortening}
	\end{table}
}

Given a superconformal primary field transforming in one of the representations of table~\ref{tab:n2-shortening}, it will be important for our analysis to know the explicit quantum numbers of all the conformal descendants.
This can be achieved efficiently by means of the Racah-Speiser algorithm \cite{Dolan:2002zh}, which has been described in great detail in \cite{Cordova:2016emh}.
Note that the weights of the supercharges in our conventions are
\begin{align}
\begin{split}\
\QQ^{\bs 1}_1 \sim \left[ + \tfrac{1}{2}, + \tfrac{1}{2}, + \tfrac{1}{2} \right], \quad
\QQ^{\bs 1}_2 \sim \left[ + \tfrac{1}{2}, - \tfrac{1}{2}, + \tfrac{1}{2} \right], \\
\QQ^{\bs 2}_1 \sim \left[ + \tfrac{1}{2}, + \tfrac{1}{2}, - \tfrac{1}{2} \right], \quad
\QQ^{\bs 2}_2 \sim \left[ + \tfrac{1}{2}, - \tfrac{1}{2}, - \tfrac{1}{2} \right].
\end{split}
\end{align}
For a long multiplet, we act on the highest weight in all possible ways with the four $\QQ$'s, so we obtain a representation of dimension
\begin{align}
\dim L = 16 (2j + 1)(2R + 1).
\end{align}
In order to construct the $A_1$ supermultiplet, we need to set $\QQ^{\bs 1}_2 = 0$, since this supercharge has the weights that correspond to the null state in table~\ref{tab:n2-shortening}.
The corresponding representation has dimension
\begin{align}
\dim A_1 = 8 (1 + j + 3 R + 4 j R).
\end{align}
In a similar way, the $A_2$ multiplet is obtained by setting $\QQ^{\bs 1}_1 \QQ^{\bs 1}_2 = 0$, and the $B_1$ multiplet by setting $\QQ^{\bs 1}_1 = 0$.
The corresponding dimensions are
\begin{align}
\dim A_2 = 8 (3R + 1), \qquad 
\dim B_1 = 8 R.
\end{align}
In this work, we will be mostly concerned with the displacement operator which has protected conformal dimension $\Delta = 2$, and transforms as a vector under rotations orthogonal to the defect.
Therefore, it must have quantum numbers $[2,1,0]$, and it has to sit at the bottom component of the short multiplet that contains it.
A careful analysis of the representation theory shows that it can only be contained in the $[A_2]_{R=0}$ multiplet~\cite{Bianchi:2018zpb}
\begin{align}
\label{eq:A2_100_multiplet}
[A_2]_{R=0} \; : \;
[1, 0, 0] \to
[ \tfrac{3}{2}, \tfrac{1}{2}, \tfrac{1}{2} ] \to
[2, 1, 0].
\end{align}
Of special relevance will be the following multiplets, some of which will appear in the OPE of two displacement multiplets
\begin{align}
\begin{alignedat}{3}
\label{eq:other_multiplets}
% Short 0 1
& [B_1]_{R=1} && :  \;
  [2, 0, 1] \to 
  [\tfrac{5}{2}, \tfrac{1}{2}, \tfrac{1}{2}] \to
  [3, 0, 0], \\[0.5ex]
% Short 1 0
& [A_1]^{j=1}_{R=0} && : \; 
  [2, 1, 0] \to 
  [\tfrac{5}{2}, \tfrac{3}{2}, \tfrac{1}{2}] \to
  [3, 2, 0], \\[0.5ex]
% Short 1/2 1/2  
& [A_1]^{j=1/2}_{R=1/2} && : \; 
  [\tfrac{5}{2}, \tfrac{1}{2}, \tfrac{1}{2}] \to
  [3, 0, 0] \oplus
  [3, 1, 0] \oplus
  [3, 1, 1] \to
  [\tfrac{7}{2}, \tfrac{1}{2}, \tfrac{1}{2}] \oplus
  [\tfrac{7}{2}, \tfrac{3}{2}, \tfrac{1}{2}] \to
  [4, 1, 0], \\[0.5ex]
% Long 00
& [L]_{R=0}^{j=0} && : \;
  [\Delta, 0, 0] \to
  [\Delta + \tfrac{1}{2}, \tfrac{1}{2}, \tfrac{1}{2}] \to
  [\Delta + 1, 1, 0] \oplus [\Delta + 1, 0, 1] \to \\
  & && \quad \to
  [\Delta + \tfrac{3}{2}, \tfrac{1}{2}, \tfrac{1}{2}] \to
  [\Delta + 2, 0,0], \\[0.5ex]
% Long 1 0
& [L]_{R=0}^{j=1} && : \;  
  [\Delta ,1,0] \to
  \left[\Delta +\tfrac{1}{2},\tfrac{1}{2},\tfrac{1}{2}\right] \oplus
  \left[\Delta +\tfrac{1}{2},\tfrac{3}{2},\tfrac{1}{2}\right] \to \\ 
  & && \quad \to
  [\Delta +1,0,0] \oplus
  [\Delta +1,1,0] \oplus
  [\Delta +1,1,1] \oplus
  [\Delta +1,2,0] \to \\
  & && \quad \to
  \left[\Delta +\tfrac{3}{2},\tfrac{1}{2},\tfrac{1}{2}\right] \oplus
  \left[\Delta +\tfrac{3}{2},\tfrac{3}{2},\tfrac{1}{2}\right] \to
  [\Delta +2,1,0].
\end{alignedat}
\end{align}
When the above long operators approach the unitarity bound, we get the following recombinations rules:
\begin{align}
\label{eq:recombination-rules}
\begin{split}
 & \lim_{\Delta \to 1} [L]_{R=0}^{j=0}
  = [A_2]_{R=0} \oplus [B_1]_{R=1}, \\
& \lim_{\Delta \to 2} [L]_{R=0}^{j=1}
  = [A_1]^{j=1}_{R=0} \oplus [A_1]^{j=1/2}_{R=1/2}.
\end{split}
\end{align}
Therefore, we can think of the $[A_2]_{R=0}$ and $[A_1]^{j=1}_{R=0}$ multiplets as the longs $[L]_{R=0}^{j=0}$ and $[L]_{R=0}^{j=1}$ at their respective unitarity bounds, and $[A_1]^{j=1/2}_{R=1/2}$ as the leftover part after the recombination of $[L]_{R=0}^{j=1}$.

As we pointed out in the introduction, our setup is closely related to the work~\cite{Liendo:2018ukf}, which considered line defects in four-dimensional $\NN = 4$ theories preserving $\osp(4^*|4)$ symmetry.
By carefully studying how our $\osp(4^*|2)$ algebra is embedded in $\osp(4^*|4)$, we can decompose the multiplets of $\NN = 4$ into their $\NN = 2$ counterparts.
The most important multiplets in the $\NN = 4$ case are $\BB_1$, which contains the diplacement operator, and $\BB_2$, which is the lowest dimension multiplet in the OPE of two diplacements.
They decompose in the following way
\begin{align}
\label{eq:decompose-N4-N2}
\begin{split}
 & \BB_1 \to [A_2]_{R=0} + 2 [B_1]_{R=1/2}, \\
 & \BB_2 \to [L]_{j=R=0}^{\Delta=2} + 2 [A_2]_{R=1/2} + 3 [B_1]_{R=1}.
\end{split}
\end{align}
Therefore, the analogous of the $\BB_1$ multiplet in our setup is $[A_2]_{R=0}$, since they both contain the displacement operator.
Moreover, the role that was played by the $\BB_2$ multiplet will be played now by $[L]_{j=R=0}^{\Delta=2}$.
With the numerical results, it will become clear that this intuition is correct.

%!TEX root = ../N2_line.tex
%%%%%%%%%%%%%%%%%%%%%%%%%%%%%%%%%%%%%%%%%%%%%

\section{Superspace}
\label{sec:superspace}

Having reviewed the symmetry algebra and its representation theory, we now proceed to construct a superspace suitable for the type of correlators we want to study. 
There are several kinds of superspaces in the literature, and which one to use usually depends on the type of multiplet being studied. Harmonic superspace is quite useful to study half-BPS multiplets, while chiral superspace is more efficient for chiral multiplets. In this work we are interested in the displacement operator, which sits in a multiplet which is neither half-BPS nor chiral, however it has the simplifying feature that its highest weight is neutral under $\su(2)_j \oplus \su(2)_R$. We therefore use the most standard superspace in which we add one fermionic coordinate for each conserved $\QQ$ supercharge. In this section we will follow closely \cite{Osborn:1998qu,Park:1999pd}.

\subsection{Basic definitions}

Since we study a $1d$ CFT which preserves the supersymmetry algebra $\osp(4^*|2)$, the superspace must have one generator $\PP$ for translations, and four generators $\QQ_\ia^\bA$ for supertranslations.
These supercharges have to satisfy the algebra
\begin{align}
\{ \QQ^{\bA}_\ia, \QQ^{\bB}_\ib \} = \ee_{\ia\ib} \ee^{\bA \bB} \PP, \qquad
[\PP, \QQ^\bA_\ia] = 0,
\end{align}
where $\bA = 1,2$ and $\ia = 1,2$.
In this section we will show how to build a superspace consistent with these commutation relations, and how to obtain the natural differential and covariant derivative.
We take the coordinates of superspace to be $z^M = (x, \, \theta^\ia_{\bA})$,
and a finite supertranslation to be implemented by the operator
\begin{align}
g(z) = g(x,\theta) = \exp \left( x \PP + \theta_{\bA}^\ia \QQ^{\bA}_\ia \right).
\end{align}
The composition of two supertranslations $g(\ee, \xi) g(z) = g(z')$ can be evaluated using the Baker-Campbell-Hausdorff formula $ e^X e^Y \approx e^{X + Y + \frac{1}{2} [X,Y] }$,
giving
\begin{align}
\label{eq:supertrans}
\begin{split}
x' & = x + \ee 
- \tfrac{1}{2} \xi \theta, \\
\theta' & = \theta + \xi\,.
\end{split}
\end{align}
Here and in what follows, we use the index-free notation introduced in appendix~\ref{apx:conventions}, where for example $\xi \theta \equiv \ee_{\ia\ib} \ee^{\bA \bB} \xi_\bA^\ia \theta_\bB^\ib = \xi_\bA^\ia \theta^\bA_\ia$.
The differential of a function in superspace is defined as
\begin{align}
\dd \equiv \dd z^M \frac{\partial}{\partial z^M} 
\quad \Rightarrow \quad
\dd f 
= \dd x \frac{\partial f}{\partial x}
+ \dd \theta_\bA^\ia \frac{\partial f}{\partial \theta_\bA^\ia}\,.
\end{align}
It will prove convenient to rewrite it in terms of the covariant derivative $D^\bA_\ia$ and the ``covariant one-form'' $e(z)$.
Looking at the differential of a supertranslation~\eqref{eq:supertrans}
\begin{align}
\begin{split}
\dd x' 
& = \dd x  - \tfrac{1}{2} \xi \dd \theta, \\
\dd \theta' & = \dd \theta,
\end{split}
\end{align}
we see that it is natural to define the one-form $e(z) \equiv \dd x + \tfrac{1}{2} \theta \dd\theta$,
which has the property $e(z') = e(z)$ for any constant supertranslation.
By rewriting the differential in terms of $e(z)$, we get
\begin{align}
 \dd = e(z) \frac{\partial}{\partial x} + \dd \theta_{\bA}^{\ia} D^{\bA}_{\ia},
\end{align}
where the covariant derivative is
\begin{align}
D^\bA_\ia 
\equiv \frac{\partial}{\partial \theta^\ia_{\bA}} 
+ \frac{1}{2}
  \theta^\bA_\ia 
  \frac{\partial}{\partial x},
\qquad
\{ D^\bA_\ia, D^\bB_\ib \} 
= \ee_{\ia\ib} \ee^{\bA \bB} \frac{\partial}{\partial x}.
\end{align}
The covariant one-form $e(z)$ will be important in the next section in order to derive the Killing equation satisfied by superconformal changes of coordinates.
The covariant derivative will be important as well, when we implement shortening conditions in superspace, see section~\ref{sec:multiplets-in-superspace}.

\subsubsection{Killing equation}

After having defined the one-form $e(z)$, we are now ready to derive the equation satisfied by a superconformal change of coordinates, which will be analogous to the conformal Killing equations in standard CFT.

A superconformal transformation is defined as a change of coordinates $z \to z'(z)$ such that $e(z)$ transforms as
\begin{align}
 e(z')^2 = \Omega^2(z) e(z)^2.
\end{align}
Under a generic change of coordinates $z \to z'(z)$, we have
\begin{align}
\begin{split}
e(z') 
& = e(z) \left( 
\frac{\partial x'}{\partial x} 
- \frac{1}{2} \frac{\partial \theta'}{\partial x} \theta'
\right)
+ \dd \theta_\bA^\ia \left(
D^\bA_\ia x' 
- \frac{1}{2} \left( D^\bA_\ia \theta' \right)  \theta'
\right).
\end{split}
\end{align}
Therefore, it is clear that the superconformal Killing equations are given by
\begin{align}
\label{eq:finite_Kill_eq}
D^\bA_\ia x' 
= \frac{1}{2}
\left( D^\bA_\ia \theta' \right) \theta', \qquad
\Omega(z) =
\frac{\partial x'}{\partial x} 
- \frac{1}{2} \frac{\partial \theta'}{\partial x} \, \theta'.
\end{align}
We will see that the usual superconformal transformations solve these constraints, but it is instructive to first expand the first equation for infinitesimal transformations $x' = x + \delta x$ and $\theta' = \theta + \delta \theta$:
\begin{align}
\label{eq:infinitesimal_Kill_eq}
D^\bA_a \left( \delta x - \tfrac{1}{2} \delta \theta \, \theta \right)
= \delta \theta^\bA_\ia.
\end{align}
In this form, it is clear that there is an infinite family of superconformal transformations.
In particular, given any function $h(z)$, we can construct a solution of the Killing equation~\eqref{eq:infinitesimal_Kill_eq} with
\begin{align}
 \delta x = h - \tfrac{1}{2} \theta (Dh), \qquad
 \delta \theta_\bA^a = D_\bA^a h.
\end{align}
It is not surprising that there is an infinite number of solutions, since this is analogous to the statement that in an ordinary one-dimensional space any change of coordinates $x' = f(x)$ is conformal.

There are three particularly simple solutions to the Killing equation~\eqref{eq:finite_Kill_eq}, which can be associated with translations, supertranslations and dilatations:
\begin{align}
\begin{alignedat}{3}
 % Trans
 & \exp(a\PP) \; : \quad 
 && x' = x + a, 
 && \theta' = \theta, \\
 % Supertrans
 & \exp(\xi \QQ) \; : \quad 
 && x' = x - \tfrac{1}{2} \xi \theta, \quad 
 && \theta' = \theta + \xi, \\
 % Dilatation
 & \exp(\lambda \DD) \; : \quad 
 && x' = \lambda x,
 && \theta' = \tfrac{1}{2} \lambda \theta.
\end{alignedat}
\end{align}
Here $a$ and $\xi$ are not necessarily infinitesimal parameters, and $\lambda$ does not need to be close to one.
In the following sections we will describe how to obtain the full set of $\osp(4^*|2)$ transformations starting from the above three.

\subsubsection{Inversion}

Inversions are special types of superconformal transformations with the property $I^2 = 1$, but such that $\det I = -1$. 
Since they belong to the disconnected component of the superconformal group, they cannot be expanded infinitesimally around the identity.
To find an inversion we must require that it squares to one and satisfies the finite Killing equation~\eqref{eq:finite_Kill_eq}.
In our superspace, such a transformation is
\begin{align}
\label{eq:inversion}
x \xrightarrow{\; I \;} x_I 
= \frac{x}{x^2 + \frac{1}{8} \theta^4}, \qquad
\theta_\bA^\ia \xrightarrow{\; I \;} (\theta_I)_{\bA}^{\ia} 
=  \frac{(\sigma_3)^\ia_{\ph \ia\ib} (x \, \theta^\ib_{\bA} - \frac{1}{2}(\theta^3)_\bA^\ib)}
{x^2 + \frac{1}{8} \theta^4},
\end{align}
where $(\sigma_3)^\ia_{\ph \ia\ib}$ denotes the components of the third Pauli matrix, and the fermionic contractions $\theta^3$ and $\theta^4$ are defined in appendix~\ref{apx:conventions}.
Using equation~\eqref{eq:finite_Kill_eq} we can find the reescaling associated with the previous inversion
\begin{align}
\Omega(z) = \frac{-1}{x^2 + \frac{1}{8} \theta^4}\,.
\end{align}
Inversions provide a simple way to generate new solutions to the Killing equation~\eqref{eq:finite_Kill_eq}. 
Imagine $\LL$ is a solution, then one can compose it with two inversions to obtain a new superconformal transformation $\LL' = I \, \LL \, I$.
Using this procedure we obtain the special superconformal transformations
\begin{align}
 \KK = I \PP I, \quad \SS = I \QQ I\,, 
 \quad \Rightarrow \quad
 e^{b\KK} = I e^{b\PP} I, \quad e^{\eta\SS} = I e^{\eta\QQ} I\,.
\end{align}
Notice that this provides a definition of the finite action of $\KK$ and $\SS$ which is not limited to infinitesimal transformations.

\subsubsection{Differential operators}

Given a solution of the infinitesimal Killing equation~\eqref{eq:infinitesimal_Kill_eq}, we can use it to build a differential operator that implements the corresponding infinitesimal transformation
\begin{align}
\LL = \delta x \, \px + \delta \theta^a_\bA \pt{\bA}{a}.
\end{align}
If we compose two transformations as $[\LL_1, \LL_2] = -\LL_3$, one can show that $\delta x_3$ and $\delta \theta_3$ still satisfy the Killing equation.
From the commutation relations of the superalgebra~\eqref{eq:osp42_superalgebra_ferm}, we see that we can obtain $\MM$ and $\RR$ by looking at the anticommutator of $\QQ$ with $\SS$, schematically
\begin{align}
 \{ \QQ, \SS \} \sim \RR + \MM + \DD.
\end{align}
In this way we can construct all the differential operators $\PP, \KK, \ldots$ of our superconformal algebra. 
However, before doing so, we need to consider a slight generalization.

In general, we are interested in the action of differential operators on superfields $\OO^{I,i}(z)$ which have a conformal dimension $\Delta$, transverse-spin index $i$, and \Rsymmetry index $I$. 
If such a field is evaluated at $z=0$, then the action of the generators simplifies
\begin{align}
\label{eq:action-diffops-origin}
 \DD \OO^{I,i}(0) 
   = \Delta \OO^{I,i}(0), \quad
 \MM_\ia^{\ph \ia \ib} \OO^{I,i}(0) 
   = \left( M_\ia^{\ph \ia \ib} \right)^i_{\ph ij} \OO^{I,j}(0), \quad
 \RR^\bA_{\ph \bA \bB} \OO^{I,i}(0) 
   = \left( R^\bA_{\ph \bA \bB} \right)^I_{\ph IJ} \OO^{J,i}(0),
\end{align}
where $M_\ia^{\ph \ia\ib}$ and $R^\bA_{\ph \bA\bB}$ form representations of the transverse-spin and \Rsymmetry subalgebras.
Demanding that the differential operators act on operators at the origin as~\eqref{eq:action-diffops-origin}, and that they act on the coordinates as described in this section, we obtain\footnote{%
Here we are abusing notation by using the same symbols for the differential operators and the generators of the superalgebra.
Moreover, as usual in this type of superspace constructions, the differential operators~\eqref{eq:diff_ops_osp42} follow the commutation relations~\eqref{eq:osp42_superalgebra_bos}-\eqref{eq:osp42_superalgebra_bosferm} with an extra minus sign, i.e. $[\LL_1, \LL_2 \} = -\LL_3$.
In principle, one would need to be careful with these extra minus signs, however for the problems we will study this will not be an issue.}
\begin{align}
\label{eq:diff_ops_osp42}
\begin{split}
% Translation
\PP 
& = \px, \\[0.5em]
% Dilatation
\DD 
& = x \px + \tfrac{1}{2} \theta^{\ia}_\bA \pt{\bA}{\ia} + \Delta, \\[0.5em]
% Special conf.
\KK 
& = \left( x^2 - \tfrac{1}{8} \theta^4 \right) \px 
+ \left( x \theta^\ia_\bA + \tfrac{1}{2} (\theta^3)^\ia_\bA \right) \pt{\bA}{\ia}
+ 2 \Delta x
+ \tfrac{1}{2} \theta_\bA^\ia \theta^\bA_\ib M_\ia^{\ph \ia\ib}
- \theta_\bA^\ia \theta^\bB_\ia R^\bA_{\ph \bA\bB}
, \\[0.5em]
% Transverse spin
\MM_\ia^{\ph \ia\ib} 
& = \theta^\ib_\bA \pt{\bA}{\ia}
- \tfrac{1}{2} \delta_\ia^{\ph \ia\ib} \theta^\ic_\bC \pt{\bC}{\ic}
+ M_\ia^{\ph \ia\ib}, \\[0.5em]
% R-symmetry
\RR^\bA_{\ph \bA\bB} 
& = \theta_{\bB}^\ia \pt{\bA}{\ia}
- \tfrac{1}{2} \delta^\bA_{\ph \bA\bB}  \theta_{\bC}^\ic \pt{\bC}{\ic}
+ R^\bA_{\ph \bA\bB} , \\[0.5em]
% Super-translation
\QQ^\bA_\ia 
& = \pt{\bA}{\ia} 
- \tfrac{1}{2} \theta_\ia^\bA \px, \\[0.5em]
% Super conformal transformation
\SS_\bA^\ia 
& = -\tfrac{1}{2} \left(x \theta_\bA^\ia + \tfrac{1}{2} (\theta^3)_\bA^\ia \right) \px
+ x \pd_{\bB}^{\ib}
- \tfrac{1}{2}
\left( 
 \theta_\bA^\ia \theta_\bB^\ib
+ 3 \theta_\bA^\ib \theta_\bB^\ia
\right) \pt{\bB}{\ib}
- \Delta \theta_\bA^\ia
- \theta_\bA^\ib M_\ib^{\ph \ib\ia}
+ 2 \theta_\bB^\ia R^\bB_{\ph \bB\bA}.
\end{split}
\end{align}
Notice also that $\{ \QQ^\bA_\ia, D^\bB_\ib \} = 0$. This standard property of the covariant derivative ensures that shortening conditions constructed with it are invariant under supersymmetry.

\subsubsection{Multiplets in superspace}
\label{sec:multiplets-in-superspace}

A generic multiplet with transverse spin $j$ and \Rsymmetry $R$ can be represented in terms of a superfield
\begin{align}
\OO^{\bA_1 \ldots \bA_{2R}}_{\ia_1 \ldots \ia_{2j}}(z) 
= \OO^{(\bA_1 \ldots \bA_{2R})}_{(\ia_1 \ldots \ia_{2j})}(z),
\end{align}
where we use $(\ia_1 \ldots \ia_m)$ to denote symmetrization of the indices.
The superspace dependence is obtained by applying a supertranslation to the superfield at the origin
\begin{align}
\label{eq:superspace-dependence}
\OO^{\bA\ldots}_{\ia\ldots}(x,\theta) & 
= \exp \big( x\PP + \theta \QQ \big) \OO^{\bA\ldots}_{\ia\ldots}(0)\,.
\end{align}
The short multiplets from table~\ref{tab:n2-shortening} can be obtained by setting the conformal dimension $\Delta$ to the appropriate value, and then imposing extra shortening conditions in terms of covariant derivatives
\begin{subequations}
\label{eq:A1A2B1_superspace}
\begin{align}
\label{eq:A1_superspace}
& A_1 \; : \;
\ee^{\ia\ib} D^{(\bA}_\ia \OO^{\bB_1) \ldots \bB_{2R}}_{\ib \ib_2 \ldots \ib_{2j}} = 0, \\[0.5ex]
\label{eq:A2_superspace}
& A_2 \; : \; 
\ee^{\ia\ib} D^{(\bA}_\ia D^{\bB}_\ib \OO^{\bC_1) \ldots \bC_{2R}} = 0, \\[0.5ex]
\label{eq:B1_superspace}
& B_1 \; : \; 
D^{(\bA}_\ia \OO^{\bB_1) \ldots \bB_{2R}} = 0.
\end{align}
\end{subequations}
It is not hard to check that the content of these shortened multiplets is in perfect agreement with the decompositions in terms of conformal primaries given by the Racah-Speiser algorithm of section~\ref{sec:shortening-conditions}.
In the rest of this section we will work out explicitly the example of the displacement multiplet $[A_2]_{R=0}$.

We start with a long scalar multiplet of conformal dimension $\Delta$, namely a  superfield that carries no transverse-spin or \Rsymmetry indices. 
In equation~\eqref{eq:other_multiplets} one can see the decomposition of this multiplet in terms of conformal primaries, which in superspace takes the form
\begin{align}
\label{eq:long_scalar_superfield}
\begin{split}
\OO(x,\theta)
& = A(x)
+ \theta_\bA^\ia B^\bA_\ia(x)
+ \theta_\bA^\ia \theta_\bB^\ib \left( 
C^{\bA\bB}_{\ia\ib}(x) + E^{\bA\bB}_{\ia\ib}(x) \right)
+ (\theta^3)_\bA^\ia F^\bA_\ia(x)
+ \theta^4 G(x)\,,
\end{split}
\end{align}
where $C^{\bA\bB}_{\ia\ib} = C^{[\bA\bB]}_{(\ia\ib)}$ and $E^{\bA\bB}_{\ia\ib} = E^{(\bA\bB)}_{[\ia\ib]}$.
Expanding equation~\eqref{eq:superspace-dependence} and comparing terms, one can obtain the explicit form of the components
\begin{align}
\begin{split}
B^\bA_\ia(x) & = \QQ^\bA_\ia A(x), \\
C^{\bA\bB}_{\ia\ib}(x) & = - \tfrac{1}{2} \QQ^{[\bA}_{(\ia} \QQ^{\bB]}_{\ib)} A(x), \\
E^{\bA\bB}_{\ia\ib}(x) & = - \tfrac{1}{2} \QQ^{(\bA}_{[\ia} \QQ^{\bB)}_{\ib]} A(x), \\
 F^{\bA}_{\ia}(x) 
 & = - \tfrac{1}{9} \Big( (\QQ^3)^\bA_\ia + \tfrac{1}{2} \QQ^\bA_\ia \PP \Big) A(x), \\
G(x) & = + \tfrac{1}{144} \left( \QQ^4 + \PP^2 \right) A(x).
\end{split}
\end{align}
Some of these terms are not annihilated by $\KK$ and therefore do not correspond to conformal primaries.
By using the commutation relations~\eqref{eq:osp42_superalgebra_bos}-\eqref{eq:osp42_superalgebra_bosferm}, we see that $A$, $B^{\bA}_{\ia}$, $C^{\bA\bB}_{\ia\ib}$ and $E^{\bA\bB}_{\ia\ib}$ are indeed primaries, but we need to take
\begin{align}
\label{eq:true-conf-desc}
F^{\text{p}}(x) = F(x) - \frac{1}{2 ( 2 \Delta + 1 )} \PP B(x), \qquad
G^{\text{p}}(x) = G(x) + \frac{1}{16 ( 2 \Delta + 1 )} \PP^2 A(x).
\end{align}
The displacement superfield $\DD(z)$ corresponds to the short multiplet $[A_2]_{R=0}$, so from table~\ref{tab:n2-shortening} and equation~\eqref{eq:A2_100_multiplet} it is clear that we need to send $\Delta \to 1$, and remove the conformal descendants $E = F^{\text{p}} = G^{\text{p}} = 0$.
We are then left with the superfield
\begin{align}
\label{eq:A2R0_superfield}
\begin{split}
\DD(x,\theta)
& = A(x)
+ \theta_\bA^\ia B^\bA_\ia(x)
+ \theta_\bA^\ia \theta_\bB^\ib C^{\bA\bB}_{\ia\ib}(x)
+ \tfrac{1}{6} (\theta^3)_\bA^\ia \px B^\bA_\ia(x)
- \tfrac{1}{48} \theta^4 \px^2 A(x).
\end{split}
\end{align}
One can obtain the same expression by making an ansatz for $\DD(z)$ of the form~\eqref{eq:long_scalar_superfield} and imposing the shortening condition~\eqref{eq:A2_superspace}
\begin{align}
\label{eq:A2R0_short}
\ee^{\ia\ib} D^{(\bA}_\ia D^{\bB)}_\ib \DD(z) = 0\,.
\end{align}
Then equation \eqref{eq:A2R0_superfield} is the most general solution to this condition, or equivalently, it implies that $E = F^{\text{p}} = G^{\text{p}} = 0$.

\subsection{Correlation functions}

Having introduced the basics of our superspace, we are now ready to construct correlation functions of long and short operators.
In general, superconformal theories have additional kinematical structures when compared to standard CFTs. A well known example is that already at the three-point level there can be non-trivial superconformal invariants \cite{Osborn:1998qu}.
We start by constructing all such invariants up to four points in section~\ref{sec:invariants}, and then compute the correlation functions for scalar long operators in section~\ref{sec:long-correlators}. We finish by specifying our results to the displacement operator multiplet in section~\ref{sec:short-correlators}.

\subsubsection{Invariants}
\label{sec:invariants}

The superconformal invariants that will form the bulding blocks of our correlators can be obtained as described in \cite{Osborn:1998qu}.
The most general case we will consider in this work is that of four points $z_1, \ldots, z_4$. 
Notice that these points can be fixed to standard values in the following way
\begin{enumerate}
  \item Fix $z = 0$ by doing a translation $\PP$ with parameter $a = - x$ followed by a supertranslation $\QQ$ with parameter $\xi = -\theta$.
  \item Fix $x = \infty$ by doing a special conformal transformation $\KK$ with parameter $b = - x_I$, and then fix $\theta = 0$ using an $\SS$ transformation of parameter $\eta = -\theta_I$.
  Here we are denoting $z_I = (x_I, \theta_I)$ the coordinates obtained from $z$ by an inversion, see equation~\eqref{eq:inversion}.
\end{enumerate}
We can combine these two types of transformations to go to a frame where two of the points are fixed to $z = 0$ and $z' = (\infty, 0)$.
For our purposes, it will be convenient to work in two different frames
\begin{align}
\begin{split}
& \fr 1 : \;
z_1, \, z_2 \; \text{unfixed}, \; z_3 = 0, \; z_4 = (\infty, 0), 
\\
& \fr 2 : \;
z_1 = 0, \; z_2 = (\infty, 0), \; z_3, \, z_4 \; \text{unfixed}.
\end{split}
\end{align}
In either frame, one can construct the invariants as the combinations of the unfixed $z_i$ which are invariant under the leftover symmetry generators $\DD$, $\MM$ and $\RR$.

Consider first the case of three points in the frame \fr 2, where the only unfixed coordinates are $z_3 = (x_3, \theta_3)$.
If there is a quantity built from $\theta_3$ which is invariant under $\MM$ and $\RR$, then it must not have any uncontracted indices. As discussed in appendix~\ref{apx:conventions}, the only such object is $(\theta_3)^4$.
On the other hand, $x_3$ is automatically invariant under $\MM$ and $\RR$, and the only independent combinations of both that is also invariant under dilatations $\DD$ is
\begin{align}
\label{eq:3pt-inv-frame}
J \big|_{\fr 2} = \frac{\theta_3^4}{x_3^2}.
\end{align}
One can invert the transformations that led to the frame \fr 2, to obtain the general expression of the three-point invariant
\begin{align}
\label{eq:3pt-inv}
J = \left( 
\frac{\theta_{12}^4}{y_{12}^2} 
+ \frac{2 \, \theta_{12} \theta_{12} \theta_{23} \theta_{23}}{y_{12} \, y_{23}}
+ \text{cycl.\,perms.} 
\right)
+ \frac{2 (\theta_{12} \theta_{23} \theta_{31}) (\theta_{12} \theta_{31} \theta_{23})}
{ y_{12} \, y_{23} \, y_{31} },
\end{align}
where $y_{ij}$ and $\theta_{ij}$ are the supertranslation invariant combinations
\begin{align}
\label{eq:susy_inv_interv}
 y_{ij} = x_i - x_j - \tfrac{1}{2} \theta_{i} \theta_{j}, \qquad
 \theta_{ij} = \theta_i - \theta_j.
\end{align}
We do not provide details on how to carry out this calculation, but one can find a similar setup in Appendix A of \cite{Kos:2018glc}.
It is worth stressing how from a very simple expression for the invariant in a certain frame~\eqref{eq:3pt-inv-frame}, we obtain a much more complicated equation in the general case~\eqref{eq:3pt-inv}.

Let us now consider the four-point case, in which one of the invariants is the standard $1d$ cross-ratio, and the remaining ones correspond to nilpotent quantities.
Unlike the three-point case, with four points there is freedom in how to choose the invariants, and we fix it by working with a basis which is simple in the frame \fr 1.
In our conventions, we take the bosonic invariant to be
\begin{align}
\label{eq:bosonic_inv_F1}
z \big|_{\fr 1} = 1 - \frac{x_2}{x_1},
\end{align}
which corresponds to the supersymmetric generalization of the standard $1d$ cross-ratio $\chi = \frac{x_{12} x_{34}}{x_{13}x_{24}}$.
From the discussion of appendix~\ref{apx:conventions}, more precisely equations~\eqref{eq:four-contract-properties} and~\eqref{eq:extra-contract-properties}, one can see that a complete basis for the nilpotent invariants is\footnote{%
We remind the reader that we are using an index-free notation for the contractions of anticommuting variables, which we describe in detail in appendix~\ref{apx:conventions}.}
\begin{align}
\label{eq:nilpotent_inv_F1}
&  I_1 \big|_{\fr 1} = \frac{\theta_1 \theta_2}{x_1}, 
&& I_2 \big|_{\fr 1} = \frac{\theta_1 \theta_1 \theta_1 \theta_1}{x_1^2}, 
&& I_3 \big|_{\fr 1} = \frac{\theta_1 \theta_1 \theta_1 \theta_2}{x_1^2}, \nonumber \\
&  I_4 \big|_{\fr 1} = \frac{\theta_1 \theta_1 \theta_2 \theta_2}{x_1^2},
&& I_5 \big|_{\fr 1} = \frac{\theta_1 \theta_2 \theta_1 \theta_2}{x_1^2},
&& I_6 \big|_{\fr 1} = \frac{\theta_1 \theta_2 \theta_2 \theta_2}{x_1^2}, \\
&  I_7 \big|_{\fr 1} = \frac{\theta_2 \theta_2 \theta_2 \theta_2}{x_1^2},
&& I_8 \big|_{\fr 1} = \frac{(\theta_1 \theta_2)^3}{x_1^3},
&& I_9 \big|_{\fr 1} = \frac{\theta_1^4 \theta_2^4}{x_1^4}. \nonumber 
\end{align}
As before, one could undo the transformation that led to the frame \fr 1, and find expressions for $I_i$ in a completely general frame.
The resulting expressions are rather involved, and we do not present them here. 
Actually, for the discussions in this paper, we will mostly need $I_i$ in the frame \fr 1, and we will only need the expressions in the frame \fr 2 to obtain the shortening conditions of equation~\eqref{eq:shorteningSSSS}.
The readers interested in this calculation can find the $I_i|_{\fr 2}$ in the attached \mathematica notebook.

In order to study crossing symmetry, we will be interested in the invariants $\tilde I_i$ obtained from $I_i$ with the replacement $z_1 \leftrightarrow z_3$.
They take simple forms when expressed in terms of the original invariants,
for example the bosonic cross-ratio becomes
\begin{align}
\tilde z = 1 - z + \frac{I_1}{2}, 
\end{align}
while the nilpotent invariants become
\begin{align}
\label{eq:invariants_14_channel}
\begin{split}
& \tilde{I}_i = I_i \quad \text{for} \quad i = 1,2,8,9, \\
& \tilde{I}_3 = I_2 - I_3,  \\
& \tilde{I}_4 = I_2 - 2 I_3 + I_4, \\
& \tilde{I}_5 = I_2 - 2 I_3 + I_5, \\
& \tilde{I}_6 = I_2 - 3 I_3 + \tfrac{3}{2} I_4 + \tfrac{3}{2} I_5 - I_6, \\
& \tilde{I}_7 = I_2 - 4 I_3 + 3 I_4 +3 I_5 -4 I_6 +I_7.
\end{split}
\end{align}

\subsubsection{Scalar long multiplets}
\label{sec:long-correlators}

We are finally ready to write our first correlators. 
In analogy with standard CFT, the building block of scalar correlators are combinations $Z_{ij}^2$ of the coordinates $z_i$ and $z_j$ such that
\begin{align}
\label{eq:invariance_Zij}
Z_{ij}^2 = \frac{(Z'_{ij})^2}{\Omega(z_i') \Omega(z_j')}.
\end{align}
Here $z_i'$ represent the coordinates obtained from $z_i$ by a superconformal transformation with conformal factor $\Omega(z)$, see equation~\eqref{eq:finite_Kill_eq}.
The combination $Z^2_{ij}$ must be built out of the supertranslation invariant intervals $y_{ij}$ and $\theta_{ij}$, defined in equation~\eqref{eq:susy_inv_interv}.
At order $x^2$, the most general combination we can build from them which transforms correctly under $\DD$, $\MM$ and $\RR$ is $y_{12}^2 + k \theta_{12}^4$.
We can fix the relative coefficient by requiring that~\eqref{eq:invariance_Zij} holds also for inversions $I$, and we find
\begin{align}
Z_{ij}^2 \equiv y_{ij}^2 + \tfrac{1}{8} \theta_{ij}^4.
\end{align}
Notice that we only defined $Z_{ij}^2$ because $|Z_{ij}| = (Z_{ij}^2)^{1/2}$ does not have a simple form in terms of $y_{ij}$ and $\theta_{ij}$. 
From the above discussion, it is clear that the two-point function of long scalar fields is
\begin{align}
\label{eq:2ptfun_longs}
\left\langle \longM_1(z_1) \longM_2(z_2) \right\rangle
= \frac{\delta_{\Delta_1, \Delta_2}}
{\left( Z_{12}^2 \right)^{\Delta_1}},
\end{align}
while the three-point function is
\begin{align}
\left\langle \longM_1(z_1) \longM_2(z_2) \longM_3(z_3) \right\rangle
= \frac{\lambda_{\OO_1 \OO_2 \OO_3}(1 + c \, J)}
 {\big( Z_{12}^2 \big)^{\frac{1}{2}(\Delta_1 + \Delta_2 - \Delta_3)}
  \big( Z_{13}^2 \big)^{\frac{1}{2}(\Delta_1 + \Delta_3 - \Delta_2)}
  \big( Z_{23}^2 \big)^{\frac{1}{2}(\Delta_2 + \Delta_3 - \Delta_1)} }.
\end{align}
This has the usual form of a three-point function, except for the presence of the three-point invariant $J$ defined in~\eqref{eq:3pt-inv}, and the free parameter $c$ that cannot be fixed by superconformal symmetry.
Finally, the four-point function of long scalar fields is
\begin{align}
\label{eq:four-pt-fun-longs}
\left\langle \longM_1(z_1) \longM_2(z_2) \longM_3(z_3) \longM_4(z_4) \right\rangle
= \frac{ F \left( I_a \right) }
{ \big( Z_{12}^2 \big)^{\frac{1}{2}( \Delta_1 + \Delta_2 )}
  \big( Z_{34}^2 \big)^{\frac{1}{2}( \Delta_3 + \Delta_4 )}}
\left( \frac{ Z_{24}^2 }{ Z_{14}^2 }\right)^{\frac{1}{2} \Delta_{12}}
\left( \frac{ Z_{14}^2 }{ Z_{13}^2 }\right)^{\frac{1}{2} \Delta_{34}}
\end{align}
where $\Delta_{ij} = \Delta_i - \Delta_j$ and $F(I_a)$ is an arbitrary function of the four-point superconformal invariants. 
We can expand $F(I_a)$ in the nilpotent basis as
\begin{align}
\label{eq:expand-F-nilpotent-inv}
F(I_a) 
= f_0(z)
+ \sum_{i = 1}^9 f_i(z) I_i,
\end{align}
where $f_0(z), \ldots, f_9(z)$ are arbitrary functions not fixed by superconformal symmetry.

\subsubsection{The displacement operator}
\label{sec:short-correlators}

Our main objective in this work is to bootstrap the four-point function of the displacement operator. This operator can be obtained as the $\Delta \to 1$ limit of a long scalar, provided that the shortening condition \eqref{eq:A2R0_short} is satisfied.

For example, the two point function of the displacement multiplet is
\begin{align}
\left\langle \dispM(z_1) \dispM(z_2) \right\rangle
= \frac{1}{ Z_{12}^2 },
\end{align}
which is compatible with the shortening condition~\eqref{eq:A2R0_short}
\begin{align}
 \ee^{\ia\ib} D^{(\bA}_{1,\ia} D^{\bB)}_{1,\ib}
 \left\langle \dispM(z_1) \dispM(z_2) \right\rangle
 = 
 \ee^{\ia\ib} D^{(\bA}_{2,\ia} D^{\bB)}_{2,\ib}
 \left\langle \dispM(z_1) \dispM(z_2) \right\rangle
 = 0.
\end{align}
Similarly, the three-point function of two displacements and one long scalar $\longM$  of dimension $\Delta$ is
\begin{align}
\label{eq:three-pt-fun-DDO}
 \left\langle \dispM(z_1) \dispM(z_2) \longM(z_3) \right\rangle
 = \frac{
    \lambda_{\dispM \dispM \longM} 
    \left( 1 -\frac{\Delta (\Delta -2)}{48}  J \right)
 }{ 
    \big( Z_{12}^2 \big)^{\frac{1}{2} (2 - \Delta)} \,
    \big( Z_{13}^2 \big)^{\frac{1}{2} \Delta} \,
    \big( Z_{23}^2 \big)^{\frac{1}{2} \Delta}
 },
\end{align}
where the coefficient $c = -\frac{1}{48} \Delta (\Delta -2)$ is fixed by the the shortening conditions at points 1 and 2. 
We could also consider the three-point function of displacement operators, in which case we set $\Delta = 1$ in equation \eqref{eq:three-pt-fun-DDO}, and the shortening condition at $z_3$ is automatically satisfied.
The previous study of the three-point functions implies the following OPE selection rule
\begin{align}
\label{eq:partial_OPE}
[A_2]_{R=0} \times [A_2]_{R=0} 
\sim 1 + [A_2]_{R=0} + \sum_{\Delta > 1} [L]^\Delta_{R=j=0} + \ldots,
\end{align}
where the $\ldots$ represent long or short multiplets such that $R,j \ne 0$. One way to complete the right-hand side of this equation would be study more general three-point functions. In section~\ref{sec:superblocks-casimir} below we will follow a different route, and derive the full OPE selection rule by solving the Casimir equations.

Finally, let us consider the four-point function of displacement multiplets, which in the frame \fr 1 takes the form
\begin{align}
\label{eq:4disp-F1}
\left\langle \dispM(z_1) \dispM(z_2) \dispM(0) \dispM(\infty,0) \right\rangle
= \frac{ F \left( I_a \right) }{ Z_{12}^{2} }.
\end{align}
In this frame it is simple to impose the shortening condition~\eqref{eq:A2R0_short} at points $z_1$ and $z_2$, leading to the constraints 
{\allowdisplaybreaks
\begin{align}
\label{eq:shorteningSSLL}
  f_2(z) & =   \frac{(z+2) (1-z) f_0'(z)}{24 z}
             - \frac{1}{48} (1-z)^2 f_0''(z), \nonumber \\
  f_3(z) & = - \frac{(1-z) f_0'(z)}{6 z}
             + \frac{(z+2) f_1(z)}{6 z}
             - \frac{1}{6} (1-z) f_1'(z), \nonumber \\
  f_4(z) & =   \frac{(1-z) f_0'(z)}{8 z}
             + \frac{f_1(z)}{4 z}
             - \frac{1}{2} (z+1) f_6(z)
             + \frac{1}{4} (1-z) z f_6'(z)
             + z f_8(z), \nonumber \\
  f_5(z) & = - \frac{f_1(z)}{2 z}, \\
  f_6(z) & = - \frac{f_0'(z)}{6 z}
             + \frac{f_1(z)}{3 z}
             - \frac{1}{6} f_1'(z), \nonumber \\
  f_7(z) & =   \frac{f_0'(z)}{12 z}
             - \frac{1}{48} f_0''(z), \nonumber \\
  f_8(z) & =   \frac{f_0'(z)}{24}
             + \frac{(5 z-12) f_0''(z)}{96}
             - \frac{(z+4) (z-1) f_0{}^{(3)}(z)}{96}
             - \frac{z (z-1)^2 f_0{}^{(4)}(z)}{192} \nonumber \\*
  & \ph =
             - \frac{f_1'(z)}{4}
             + \frac{(1-z) f_1''(z)}{8}
             + 12 z f_9(z) \nonumber.
\end{align}
}%
One should also impose shortening at the points $z_3$ and $z_4$. 
The simplest way to achieve this is to consider the four-point function in the frame \fr 2, but now special care is needed since equations~\eqref{eq:bosonic_inv_F1}-\eqref{eq:nilpotent_inv_F1} are no longer valid in this frame.
All in all, one obtains one extra constraint
\begin{align}
\label{eq:shorteningSSSS}
\begin{split}
 f_9(z) = 
 & - \frac{\left(z^2+z+2\right) f_0'(z)}{288 z^3}
   + \frac{(z (4-5 z)+8) f_0''(z)}{1152 z^2}
   + \frac{(z+4) (z-1) f_0{}^{(3)}(z)}{1152 z} \\
 & + \frac{(z-1)^2 f_0{}^{(4)}(z)}{2304}
   - \frac{(z+2) f_1(z)}{144 z^3}
   + \frac{(z+2) f_1'(z)}{144 z^2}
   + \frac{(z-1) f_1''(z)}{144 z}.
\end{split}
\end{align}
Summarizing, we have found that the four-point function of displacements depends on two unfixed functions $f_0(z)$ and $f_1(z)$.
These two functions will be the subject of the bootstrap analysis of the following sections.

%!TEX root = ../N2_line.tex
%%%%%%%%%%%%%%%%%%%%%%%%%%%%%%%%%%%%%%%%%%%%%

\section{Superconformal blocks}
\label{sec:superblocks}

Armed with the four-point functions in superspace we can now calculate the relevant superconformal blocks. 
There are several approaches that have been used to calculate superblocks with varying degrees of success. 
These include explicit calculation of three-point couplings of descendants~\cite{Poland:2010wg,Fortin:2011nq}, the shadow formalism \cite{Fitzpatrick:2014oza,Li:2018mdl}, Ward identities in harmonic superspace \cite{Dolan:2004mu,Doobary:2015gia,Liendo:2015cgi,Lemos:2016xke}, the Casimir operator \cite{Fitzpatrick:2014oza,Bissi:2015qoa,Cornagliotto:2017dup,Ramirez:2018lpd}, and the connection to Calogero-Sutherland models \cite{Buric:2019rms}. 
Because the multiplets we are considering are scalars with no \Rsymmetry or transverse-spin indices, we will use the most conventional of these methods, which is to consider superblocks as eigenfunctions of the Casimir operator.\footnote{In some selected cases we will also calculate three-point couplings of descendants as a non-trivial check for our computations.}
In the main text we will concentrate on the blocks for the displacement multiplet, however in appendix~\ref{apx:long-blocks} we present more general correlators that also include non-protected long operators.

\subsection{From the Casimir equation}
\label{sec:superblocks-casimir}

Superconformal blocks are given by a finite sum of $1d$ bosonic blocks, that capture the contributions of the $\sl(2;\mathbb R)$ primaries in the conformal multiplets:
\begin{align}
\label{eq:chiral-block}
g_{\Delta}^{\Delta_{12},\Delta_{34}}(z)
= z^\Delta {}_2F_1(\Delta - \Delta_{12}, \Delta + \Delta_{34}, 2\Delta, z).
\end{align}
The coefficients in this sum are fixed by supersymmetry, so we can make an ansatz for the functions $f_i$ in terms of bosonic blocks. After acting with the Casimir operator on the four-point function, we will obtain a coupled system of equations for the functions $f_i$ that we will use to fix the coefficients in our ansatz.
Since we will use the coupled set of differential equations only to fix these coefficients, the superblocks will automatically satisfy the correct boundary conditions.

The Casimir of the $\osp(4^*|2)$ superalgebra is given by
\begin{align}
\label{eq:casimir_def}
\begin{split}
\CC^2 =
& + \DD^2 
- \tfrac{1}{2} ( \PP \KK + \KK \PP )
+ \tfrac{1}{2} \MM_\ia^{\ph \ia\ib} \MM_\ib^{\ph \ib\ia}
- \RR^\bA_{\ph \bA\bB} \RR^\bB_{\ph \bB\bA}
- \tfrac{1}{2} [ \QQ^\bA_\ia, \SS_\bA^\ia ].
\end{split}
\end{align}
When it acts on an operator $\OO$ with quantum numbers $[\Delta, j, R]$ it has the following eigenvalue
\begin{align}
\CC^2 \, \OO = \mathfrak c_{\Delta,j,R} \, \OO, \qquad
\mathfrak c_{\Delta,j,R} = \Delta(\Delta + 1) + j(j+1) -2 R(R+1).
\end{align}
Given a four-point function, we can evaluate it by taking OPEs in the $(12)\to(34)$ channel, leading to the usual expansion in terms of superconformal blocks
\begin{align}
\label{eq:expansion-superconf-blocks}
 \langle \dispM(z_1) \dispM(z_2) \dispM(z_3) \dispM(z_4) \rangle 
 = \frac{1}{Z_{12}^2 Z_{34}^2} \sum_{\OO \in \DD \times \DD} \lambda^2_{\DD\DD\OO} \, \GG_{\OO}(I_a).
\end{align}
In order to obtain a superconformal block, we act with the Casimir on the four-point function and find the solution to the eigenvalue problem\footnote{Notice that the dependence on $Z_{12}^2$ drops from the eigenvalue problem since $\CC_{12}^2 Z_{12}^2 = 0$.}
\begin{align}
 \CC^2_{12} \, \GG_{\Delta,j,R}(I_a) = \mathfrak c_{\Delta,j,R} \, \GG_{\Delta,j,R}(I_a).
\end{align}
The differential operator $\CC^2_{12}$ is constructed from the Casimir~\eqref{eq:casimir_def} and the symmetry generators in differential form~\eqref{eq:diff_ops_osp42}.
Note that the operators need to be evaluated at points $z_1$ and $z_2$, namely $\LL_{12} = \LL_1 + \LL_2$.
In order to solve the above equation, we take $\GG$ to be of the form~\eqref{eq:expand-F-nilpotent-inv} with the shortening conditions~\eqref{eq:shorteningSSLL} and~\eqref{eq:shorteningSSSS}.
Furthermore, we evaluate the Casimir equation in the frame \fr 1 where the calculations are simpler. The resulting system of differential equations is
\begin{subequations}
\label{eq:casimirEq_SSSS}
\begin{align}
 \label{eq:casimirEq_SSSS_a}
 & -z^2 \big[ (z-1) f_0''(z) + f_0'(z) \big]-4 z f_1(z) 
  = \mathfrak c_{\Delta,j,R} \, f_0(z), \\
 \label{eq:casimirEq_SSSS_b}
 & -(z-1) z \big( z f_1''(z) + 4 f_1'(z) \big) +(2-z) \big( \tfrac{1}{2} f_0'(z)+2 f_1(z) \big)
  = \mathfrak c_{\Delta,j,R} \, f_1(z).
\end{align}
\end{subequations}
Notice the similarity of~\eqref{eq:casimirEq_SSSS_a} with the usual non-supersymmetric $1d$ Casimir equation.
To solve these equations one should make an ansatz for the $f_i$ in terms of $1d$ bosonic blocks.
However, as discussed in~\cite{Cornagliotto:2017dup}, it is simpler to first ``change basis'' to a set of functions $G_i(z)$, where each of the $G_i$ captures the contribution of the external superconformal descendants, and build an ansatz for the $G_i$ instead.
Let us review in detail how to implement this idea.

We start by expanding the displacement multiplets in terms of their conformal descendants~\eqref{eq:A2R0_superfield}, so that the four-point function becomes
\begin{align}
\label{eq:expansionDDDD_confPrim}
\begin{split}
 \left\langle \dispM(z_1) \dispM(z_2) \dispM(0) \dispM(\infty,0) \right\rangle
& = \langle A(x_1) A(x_2) A(0) A(\infty) \rangle \\
& - \theta_{1,\bA}^\ia \theta_{2,\bB}^\ib 
    \langle B_{\ia}^\bA(x_1) B_{\ib}^\bB(x_2) A(0) A(\infty) \rangle 
 + \ldots
\end{split}
\end{align}
Note that since we work in the frame \fr 1, we have $\theta_3 = \theta_4 = 0$, so only the superconformal primary $A$ at points $3$ and $4$ will appear.
There are only three four-point functions of descendants that contribute to the above expansion, and for each of them we define a new function $G_i$ as
\begin{align}
\label{eq:ansatz_change_basis_SSLL}
\begin{alignedat}{3}
&  \langle A(x_1) A(x_2) A(0) A(\infty) \rangle 
&& \to \;\;
&& \frac{1}{|x_{12}|^2} \, G_0(z), \\
&  \langle B_{\ia}^\bA(x_1) B_{\ib}^\bB(x_2) A(0) A(\infty) \rangle 
&& \to \;\;
&& \frac{x_{12} \, \ee^{\bA\bB} \ee_{\ia \ib}}{|x_{12}|^4} \, G_1(z), \\
&  \langle C_{\ia\ib}^{\bA\bB}(x_1) C_{\ic\id}^{\bC\bD}(x_2) A(0) A(\infty) \rangle 
&& \to \;\;
&& \frac{\ee^{\bA\bB}\ee^{\bC\bD} (\ee_{\ia \ic} \ee_{\ib \id} + \ee_{\ia \id} \ee_{\ib \ic})}{|x_{12}|^4} G_2(z), \\
\end{alignedat}
\end{align}
On one hand, we can introduce~\eqref{eq:ansatz_change_basis_SSLL} in the expansion~\eqref{eq:expansionDDDD_confPrim}, and on the other, we can expand the four-point function of displacements~\eqref{eq:4disp-F1} in terms of $\theta_1$ and $\theta_2$.
By matching the components of the two sides, we get that the change of basis is
\begin{align}
\label{eq:change_basis_SSSS}
 f_0(z) = G_0(z), \qquad
 f_1(z) = - \tfrac{1}{z} \big[ G_0(z) + G_1(z) \big].
\end{align}
Furthermore, we see that $G_2$ must be related to $G_0$ and $G_1$ by
\begin{align}
& G_2(z) = 
    \tfrac{1}{8} G_0(z)
  + \tfrac{1}{48} z (z-4) G_0'(z)
  - \tfrac{1}{48} z^2 (z-1) G_0''(z)
  + \tfrac{1}{2} G_1(z)
  + \tfrac{1}{12} z (z-2) G_1'(z).
\end{align}
It is natural that $G_2$ is related to $G_0$ and $G_1$, since the four-point function of displacements contains only two unfixed functions $f_0(z)$ and $f_1(z)$. However, we still had to include $G_2$ in~\eqref{eq:ansatz_change_basis_SSLL}, because a priori we did not know what this relation was.

The virtue of the $G_i$ basis is that now the ansatz in terms of $1d$ bosonic blocks is very simple
\begin{align}
\label{eq:ansatz-sum-bosonic-blocks}
 G_i(z)
 = a_i \, g_{\Delta}^{0,0}(z)
 + b_i \, g_{\Delta+\tfrac{1}{2}}^{0,0}(z)
 + c_i \, g_{\Delta+1}^{0,0}(z)
 + d_i \, g_{\Delta+\tfrac{3}{2}}^{0,0}(z)
 + e_i \, g_{\Delta+2}^{0,0}(z).
\end{align}
We finally have all the ingredients to solve the Casimir equations~\eqref{eq:casimirEq_SSSS}.
If we consider the case of an exchanged multiplet $[\Delta, 0, 0]$, then the Casimir eigenvalue is $\cc = \Delta(\Delta+1)$, and the equations are solved by
\begin{align}
\label{eq:sol_SSSS_00}
\begin{split}
& G_0(z) 
  = g_\Delta^{0,0}(z)
  + \frac{(\Delta -1) \Delta  (\Delta +1)}{4 (\Delta +2) (2 \Delta +1) (2 \Delta +3)}
    g_{\Delta+2}^{0,0}(z), \\[0.5em]
& G_1(z) 
  = \frac{1}{2} (\Delta - 2) g_\Delta^{0,0}(z)
  -\frac{(\Delta -1) \Delta  (\Delta +1) (\Delta +3)}{8 (\Delta +2) (2 \Delta +1) (2 \Delta +3)}
  g_{\Delta+2}^{0,0}(z).
\end{split}
\end{align}
From now on, we will sometimes use vectorial notation $G(z) = \left( G_0(z), G_1(z) \right)$.
Depending on the value of $\Delta$, the solution~\eqref{eq:sol_SSSS_00} is interpreted as follows:
\begin{itemize}
 \item For $\Delta = 0$ the block reduces to $G_{\mathds 1}(z) = (1, -1)$, and corresponds to the identity operator being exchanged.
 \item For $\Delta = 1$ the block reduces to $G_{A_2}(z) = \big(g_{1}^{0,0}(z), -\tfrac{1}{2} g_{1}^{0,0}(z) \big) $, and corresponds to a displacement multiplet $[A_2]_{R=0}$ being exchanged.
 \item For $\Delta > 1$ the block $G_\Delta^{[0,0]}(z)$ is given by~\eqref{eq:sol_SSSS_00}, and corresponds to a long scalar multiplet $[L]^{j=R=0}_{\Delta}$ being exchanged. 
\end{itemize}
One can also consider an exchanged multiplet $[\Delta, 1, 0]$, in which case the Casimir eigenvalue is $\cc = \Delta(\Delta+1) + 2$, and the equations are solved by
\begin{align}
\label{eq:sol_SSSS_10}
G_0(z) = g_{\Delta+1}^{0,0}(z), \qquad
G_1(z) = -\tfrac{1}{2} g_{\Delta+1}^{0,0}(z).
\end{align}
The solution~\eqref{eq:sol_SSSS_10} is interpreted as follows:
\begin{itemize}
  \item For $\Delta = 2$ the block reduces to $G_{A_1}(z) = \big(g_{3}^{0,0}(z), -\tfrac{1}{2} g_{3}^{0,0}(z) \big)$. Note that from the recombination rules~\eqref{eq:recombination-rules}, we could interpret the solution as either an $[A_1]^{j=1}_{R=0}$ or an $[A_1]^{j=1/2}_{R=1/2}$.
   The correct interpretation is that it is actually $[A_1]^{j=1/2}_{R=1/2}$ which is exchanged, in particular its descendant with quantum numbers $[3,0,0]$, see equation~\eqref{eq:other_multiplets}.
  \item For $\Delta > 2$ the block $G_\Delta^{[1,0]}(z)$ is given by~\eqref{eq:sol_SSSS_10}, and corresponds to a long scalar multiplet $[L]^{j=1, R=0}_{\Delta}$ being exchanged. 
\end{itemize}
We have tried solving the Casimir equation considering other possible exchanges, but in all cases there were no new solutions found, so the above are all the operators that can appear in the OPE of two displacement multiplets.

\paragraph{OPE selection rule.} Summarizing the above results, we obtain the following selection rule
\begin{align}
\label{eq:selection-rule-DDO}
[A_2]_{R=0} \times [A_2]_{R=0} \sim
1 + 
[A_2]_{R=0} +
[A_1]_{R=1/2}^{j=1/2} +
\sum_{\Delta > 1} [L]_\Delta^{[0,0]} +
\sum_{\Delta > 2} [L]_\Delta^{[1,0]},
\end{align}
which completes the partial selection rule \eqref{eq:partial_OPE} obtained from the three-point function analysis.

\subsection{From two- and three-point functions}
\label{sec:blocks-two-three-pt-funs}

In this section, we calculate the superconformal blocks in the $[\Delta, 0, 0]$ channel~\eqref{eq:sol_SSSS_00} following the approach of~\cite{Poland:2010wg}.
This provides a non-trivial consistency check for our results, and sheds light on the structure of such blocks.
The key insight is that the coefficients appearing in the superconformal blocks are OPE coefficients and norms of conformal descendants
\begin{align}
\label{eq:blocks00_as_opecoef}
\begin{split}
 & G_0(z) 
 = \frac{\lambda_{AAA}^2}{\langle A | A \rangle} \, g_{\Delta    }^{0,0}(z)
 + \frac{\lambda_{AAG}^2}{\langle G | G \rangle} \, g_{\Delta + 2}^{0,0}(z), \\
 & G_1(z) 
 = \frac{\lambda_{AAA} \lambda_{BBA}}{\langle A | A \rangle} \, g_{\Delta    }^{0,0}(z)
 + \frac{\lambda_{AAG} \lambda_{BBG}}{\langle G | G \rangle} \, g_{\Delta + 2}^{0,0}(z).
\end{split}
\end{align}
Here $\lambda_{O_1 O_2 O_3}$ denotes the OPE coefficient of two fields from the displacement multiplet with one operator from a long scalar multiplet, namely  $O_1, O_2 \in \DD$ and $O_3 \in \OO$, see equations~\eqref{eq:long_scalar_superfield} and~\eqref{eq:A2R0_superfield} for more details.
On the other hand, $\langle O | O \rangle$ denotes the norm of an operator that belongs to the long multiplet $\OO$, and can be computed from the two-point function as explained below.

The procedure to obtain the OPE coefficients resembles the way we obtained the change of basis in equation~\eqref{eq:change_basis_SSSS}.
Let us take the three-point function~\eqref{eq:three-pt-fun-DDO} of two displacement operators and a long scalar of dimension $\Delta$.
On one hand, we expand it in the fermionic variables, while on the other we expand the external superfields in terms of their conformal descendants~\eqref{eq:long_scalar_superfield} and~\eqref{eq:A2R0_superfield}
\begin{align}
\begin{alignedat}{2}
  \langle \DD(z_1) \DD(z_2) \OO(z_3) \rangle 
&  = \frac{\lambda_{\DD \DD \OO}}{|x_{12}|^{2-\Delta} |x_{13}|^\Delta |x_{23}|^\Delta}
&& - \theta_{1,\bA}^\ia \theta_{2,\bB}^\ib 
     \frac{\tfrac{1}{2}(\Delta-2) \lambda_{\DD \DD \OO} \ee_{\ia\ib}\ee^{\bA\bB}}
          {|x_{12}|^{3-\Delta} |x_{13}|^\Delta |x_{23}|^\Delta}
   + \ldots \\[0.5em]
&  = \langle A(x_1) A(x_2) A(x_3) \rangle 
&& - \theta_{1,\bA}^\ia \theta_{2,\bB}^\ib \langle B^\bA_\ia(x_1) B^\bB_\ib(x_2) A(x_3) \rangle
   + \ldots
\end{alignedat}
\end{align}
Mapping the two sides one can obtain all the OPE coefficients of the descendant fields.
The relevant ones for us will be
\begin{align}
\label{eq:DDL_opecoefs}
\begin{alignedat}{2}
&  \lambda_{AAA}
   = \lambda _{\DD \DD \OO},
&& \lambda_{AAG}
   = - \frac{(\Delta -1) \Delta  (\Delta +1) \lambda _{\DD\DD\OO}}{24 (2 \Delta +1)}, \\
&  \lambda _{BBA}
   = \frac{1}{2} (\Delta -2) \lambda _{\DD\DD\OO},
&& \lambda _{BBG}
   = \frac{(\Delta -1) \Delta  (\Delta +1) (\Delta +3) \lambda _{\DD\DD\OO}}{48 (2 \Delta +1)}, \\
&  \lambda_{CCA}
   = - \frac{1}{16} (\Delta -3) (\Delta -2) \lambda _{\DD\DD\OO}, \quad \;
&& \lambda_{CCG}
   = \frac{(\Delta -1) \Delta  (\Delta +1) (\Delta +3) (\Delta +4) \lambda _{\DD\DD\OO}}{384 (2 \Delta +1)}.
\end{alignedat}
\end{align}
Notice how $ \lambda_{A A G} $, $\lambda _{B B G}$, $ \lambda_{C C G}$ vanish for $\Delta = 1$, as expected from the shortening $\OO \to \DD$ and the fact that $G \notin \DD$.
We can do a similar analysis for the two-point function~\eqref{eq:2ptfun_longs} of scalar longs of dimension $\Delta$. 
In this case we obtain the norms of the descendants
\begin{align}
\label{eq:long_norms}
\begin{alignedat}{2}
&  \langle A | A \rangle 
   = 1, 
&& \langle E | E \rangle 
   = \frac{1}{8} (\Delta -1) \Delta , \\
&  \langle B | B \rangle 
   = \Delta , 
&& \langle F | F \rangle 
   = \frac{2 (\Delta -1) \Delta  (\Delta +1) (\Delta +2)}{9 (2 \Delta +1)}, \\
&  \langle C | C \rangle 
   = \frac{1}{8} \Delta  (\Delta +2), \qquad \quad
&& \langle G | G \rangle
   = \frac{(\Delta -1) \Delta  (\Delta +1) (\Delta +2) (2 \Delta +3)}{144 (2 \Delta +1)}.
\end{alignedat}
\end{align}
It is a simple exercise to check that inserting~\eqref{eq:DDL_opecoefs} and~\eqref{eq:long_norms} in~\eqref{eq:blocks00_as_opecoef} leads to the superconformal blocks~\eqref{eq:sol_SSSS_00}.
One could do a similar analysis to compute the blocks in the $[\Delta, 1, 0]$ channel, but it would be more involved, since then an expression for the three-point functions of external operators with transverse spin would be needed.

\subsection{Crossing equations}
\label{sec:crossing-equations}

In the previous sections we have studied the four-point function of displacement operators in the $(12)\to(34)$ channel.
Demanding that it is equivalent to the four-point function in the $(14)\to(23)$ channel leads to the crossing equation
\begin{align}
\frac{1}{Z_{12}^2 Z_{34}^2} \left(
  f_0(z) + \sum_{i=1}^9 I_i f_i(z)
\right)
= \frac{1}{Z_{14}^2 Z_{23}^2}
\left(
f_0(\tilde z)
+ \sum_{i=1}^9 \tilde I_i f_i(\tilde z)
\right),
\end{align}
where the $\tilde I_i$ invariants appear in equation~\eqref{eq:invariants_14_channel}, and are obtained from the $I_i$ by the replacement $z_1 \leftrightarrow z_3$.
Since $\tilde z = 1 - z + \frac{1}{2} I_1$, we can Taylor expand the $f_i$'s in the right-hand side around $\tilde z = 1 - z$, and insert the expressions for the $\tilde I_i$.
By looking at independent terms, one can see that the crossing equation reduces to
\begin{align}
\label{eq:crossing-equation}
 (1-z)^2 H(z) - z^2 H(1-z) = 0,
\end{align}
where $H(z)$ is a two-dimensional vector with components
\begin{align}
\label{eq:define-H}
\begin{split}
& H_0(z) = G_0(z), \\
& H_1(z) = - 2 z G_0(z) + z (z-1) G_0'(z) - 4(z-1) G_1(z).
\end{split}
\end{align}
Notice that from the first component we obtain the usual $1d$ bosonic crossing equation, but the second mixes $G_0(z)$ and $G_1(z)$ in a non-trivial way.

\subsection{An exact solution}

In this section we present a family of exact solutions to the crossing equations in terms of free fields.
We will argue in section~\ref{sec:analytical} that one solution in this family describes the strong coupling limit of line defects that admit a holographic description. 
Furthermore, these solutions will play a prominent role in the next two sections, where we will apply numerical and analytical bootstrap techniques to this correlator. 

The most general solution of crossing that we have found built from Wick contractions contains one free parameter $\xi$.
Since it is a valid correlator, it can be expanded in terms of superconformal blocks as in equation~\eqref{eq:expansion-superconf-blocks}
\begin{align}
\label{eq:free-field-theory}
\begin{split}
\langle \DD(z_1) \DD(z_2) \DD(z_3) \DD(z_4) \rangle
&= \frac{1}{Z_{12}^2 Z_{34}^2 }
\left[ 
1
+ \xi \frac{Z_{12}^2 Z_{34}^2}{Z_{13}^2 Z_{24}^2 }
+     \frac{Z_{12}^2 Z_{34}^2}{Z_{14}^2 Z_{23}^2 }
\right] \\
&= \frac{1}{Z_{12}^2 Z_{34}^2 }
\left[ 
1
+ c \, \GG_{A_1}
+ \sum_{\Delta \ge 2} a_\Delta \GG_{L_\Delta^{[0,0]}}
+ \sum_{\Delta \ge 3} b_\Delta \GG_{L_\Delta^{[1,0]}}
\right].
\end{split}
\end{align}
Notice how the block $\GG_{A_2}$, which a priori could appear in the expansion, has vanishing OPE coefficient $\lambda_{A_2}^2 = 0$ for any value of $\xi$.
The other OPE coefficients are given by
\begin{align}
\label{eq:ope-coefs-free-theory}
 a_\Delta 
  = \frac{ \left(1 + (-1)^{\Delta } \xi\right) \sqrt{\pi} \, \Gamma(\Delta +3)}{2^{2 \Delta +1} \Gamma \left(\Delta +\frac{1}{2}\right)}, \qquad
 b_\Delta = \frac{3 (\Delta -1) }{2 (\Delta +1)} \frac{ \left(1 + (-1)^{\Delta +1} \xi\right) \sqrt{\pi} \, \Gamma(\Delta +3)}{2^{2 \Delta +1} \Gamma \left(\Delta +\frac{1}{2}\right)},
\end{align}
and $ c = b_{\Delta = 2} = (1-\xi)/2$.
Positivity of the OPE coefficients requires $-1 \le \xi \le 1$. The theory with $\xi = 1$ corresponds to free bosons, $\xi = -1$ corresponds to free fermions, and certain values $-1 < \xi < 1$ correspond to free gauge theories, as discussed in~\cite{Liendo:2018ukf}.
We will argue in section~\ref{sec:analytical} that the bosonic theory with $\xi = 1$ corresponds to the displacement operator at leading order in the strong-coupling limit. The physical interpretation of the fermionic $\xi = -1$ theory is less clear, since we know that the displacement must be a bosonic operator. Nevertheless, it will be important as a valid solution of crossing that will sit in interesting corners of the allowed regions of the numerical results in next section.

%!TEX root = ../N2_line.tex
%%%%%%%%%%%%%%%%%%%%%%%%%%%%%%%%%%%%%%%%%%%%%

\section{Numerical results}
\label{sec:numerical}

In this section we use numerical boostrap techniques~\cite{Rattazzi:2008pe,Poland:2011ey,Kos:2014bka} to bound conformal dimensions and OPE coefficients of operators that appear in the four-point function of displacement operators.
We start each subsection with a short review of the numerical algorithm, and then we proceed to discuss the results. 
We have generated tables of derivatives of superconformal blocks with \mathematica, which are then used by the semidefinite program solver \texttt{SDPB}~\cite{Simmons-Duffin:2015qma}\footnote{%
An alternative to \mathematica to compute the tables is \texttt{PyCFTBoot}~\cite{Behan:2016dtz}, which then relies on \texttt{SDPB} to carry out the optimizations.
On the other hand, one can generate the tables in \mathematica, but then perform the numerics in \texttt{JuliBoots}~\cite{Paulos:2014vya}.
}.
The results are analyzed using \texttt{python}, and the plots are generated with \texttt{matplotlib}~\cite{Hunter:2007}.

In section~\ref{sec:crossing-equations} we derived the crossing equations~\eqref{eq:crossing-equation}, which take the simple form $F(z) = 0$ in terms of the two-dimensional vector 
\begin{align}
F(z) \equiv (1-z)^2 H(z) - z^2 H(1-z) \, .
\end{align}
We can expand $F(z)$ summing the contributions of the operators that appear in the OPE of two displacements~\eqref{eq:selection-rule-DDO}
\begin{align}
\label{eq:crossing-no-assumtions}
 F(z) =
 F_{\mathds 1}(z) + 
 \lambda_{A_1}^2 F_{A_1}(z) + 
 \lambda_{A_2}^2 F_{A_2}(z) + 
 \sum_{\Delta > 1} \lambda^2_{L_\Delta^{[0,0]}} F_\Delta^{[0,0]}(z) + 
 \sum_{\Delta > 2} \lambda^2_{L_\Delta^{[1,0]}} F_\Delta^{[1,0]}(z)
 = 0 \, ,
\end{align}
where by unitarity the OPE coefficients are real, hence $\lambda^2_\OO \ge 0$.
Here and in what follows we are using a shorthand notation where it is implicitly understood that $\lambda^2_\OO = \lambda^2_{\DD\DD\OO}$.

In order to explore the numerical constraints implied by crossing we will make some structural assumptions about the CFT data.
In some of our plots we will assume that $\lambda^2_{A_2} = 0$, or equivalently, that the displacement multiplet does not appear in the OPE of two displacements. Notice that this is the case for the mean-field solutions of the previous section, as well as for $\Nm = 4$ theories that are interpreted as $\Nm = 2$ SCFT \cite{Giombi:2017cqn}.
This is also true whenever the displacement is odd under a $\mathbb{Z}_2$ symmetry.
One could relax this condition, however we found that the numerical results become significantly weaker. It will be interesting to explore this further in the future. 
The second assumption is that the low-lying spectrum is somehow sparse, with gaps in between the local operators. More precisely, we will assume an isolated long operator  with dimension $\Delta_{[0,0]}$ separated by a finite gap from the unitarity bound, and a second gap between $\Delta_{[0,0]}$ and a continuum of long operators with dimensions $\Delta \ge \Delta_{[0,0]}'$. Similar assumptions will also be made for the longs in the $[1,0]$ channel.

The most general case we will be studying is then
\begin{align}
\label{eq:crossing-with-assumtions}
\begin{split}
    F_{\mathds 1}(z)
  + \lambda_{A_1}^2 F_{A_1}(z)
  + \lambda_{A_2}^2 F_{A_2}(z)
& + \lambda_{L_{\Delta_{[0,0]}}}^2 F_{\Delta_{[0,0]}}^{[0,0]}(z)
  + \lambda_{L_{\Delta_{[1,0]}}}^2 F_{\Delta_{[1,0]}}^{[1,0]}(z) \, + \\
& + \sum_{\Delta \ge \Delta'_{[0,0]}} \lambda^{2}_{L_\Delta^{[0,0]}} F_\Delta^{[0,0]}(z) 
  + \sum_{\Delta \ge \Delta'_{[1,0]}} \lambda^{2}_{L_\Delta^{[1,0]}} F_\Delta^{[1,0]}(z)
  = 0 \, .
\end{split}
\end{align}

When we discuss the results, it will be instructive to compare with the free-field solutions~\eqref{eq:free-field-theory}.
In the plots we will represent these solutions with a solid bullet $\bullet$ or dashed line \tikz[baseline]{\draw[dotted] (0,.5ex)--++(.5,0) ;}, accompanied by a letter representing the type of solution
\begin{align}
\label{eq:free-solutions-bullets}
\begin{split}
 & \bullet \, B: \; \text{Free boson}, \; \xi = 1, \\
 & \bullet \, F: \; \text{Free fermion}, \; \xi = -1, \\
 & \bullet \, G: \; \text{Free gauge theory}, \; -1 < \xi < 1.
\end{split}
\end{align}
Currently, the only $\Nm = 2$ line defect with insertions that has been studied in the literature is the one in $\Nm = 4$ SYM. The leading-order correlation function of $\Dm$'s at strong coupling was computed in~\cite{Giombi:2017cqn}, and it is given by the free bosonic solution. In that work, the first-order correction in $\frac{1}{\sqrt{\lambda}}$ to the correlator was also obtained.
It would be an interesting problem for the future to study an $\Nm = 2$ line defect with insertions either using holography or perturbation theory and compare with our numerical bounds. 

\subsection{Dimension bounds}

The algorithm for bounding operator dimensions works in the following way.
First, one assumes a spectrum of operator dimensions.
In the case of interest to us~\eqref{eq:crossing-with-assumtions}, this boils down to fixing the dimension of the isolated longs $\Delta_{[0,0]}$ and $\Delta_{[1,0]}$, and also the dimension of the first longs in the continuum $\Delta'_{[0,0]}$ and $\Delta'_{[1,0]}$.
Then one tries to find a functional $\alpha$ such that
\begin{align}
\label{eq:bound-dim-1}
\alpha(F_{\mathds 1}) = 1 \, , \quad 
\alpha(F_{\II}) \ge 0 \, , \quad
\alpha \big(F_\Delta^{[0,0]} \big) \ge 0 \;\; \text{for} \;\; \Delta \ge \Delta_{[0,0]}' \, , \quad
\alpha \big(F_\Delta^{[1,0]} \big) \ge 0 \;\; \text{for} \;\; \Delta \ge \Delta_{[1,0]}' \, ,
\end{align}
where $\II = A_1,\, A_2,\, L^{[0,0]}_{\Delta_{[0,0]}},\, L^{[1,0]}_{\Delta_{[1,0]}}$ runs over all the operators with fixed conformal dimensions.
If such functional $\alpha$ exists, then it is not possible to satisfy equation~\eqref{eq:crossing-with-assumtions}, and the spectrum is ruled out.

As is customary we consider functionals of the form
\begin{align}
\label{eq:approx-conf-blocks}
 \alpha(F_\Delta) 
 = \sum_{i=0}^1 \sum_{m=0}^\Lambda a_{i,m} 
   \left. \frac{\partial^m F_{i,\Delta}(z)}{\partial z^m} \right|_{z=1/2}
 \approx \chi(\Delta) P(\Delta) \, ,
\end{align}
where $i = 0,1$ runs over the two components of $F_\Delta(z)$, and the number of derivatives $\Lambda$ needs to be increased in order to obtain stronger bounds.
In the last step we have approximated the conformal blocks by a positive function $\chi(\Delta) \ge 0$ multiplying a linear combination of polynomials in $\Delta$
\begin{align}
\label{eq:poly-approx-conf-blocks}
 P(\Delta) = \sum_{i=0}^1 \sum_{m=0}^\Lambda a_{i,m} P_{i,m}(\Delta) \, .
\end{align}
This approximation can be obtained as described in~\cite{Kos:2013tga,Kos:2014bka}.
Thanks to~\eqref{eq:approx-conf-blocks} and~\eqref{eq:poly-approx-conf-blocks}, we can reformulate the optimization problem~\eqref{eq:bound-dim-1} as finding a set of coefficients $a_{i,m}$ such that
\begin{align}
\label{eq:bound-dim-2}
\begin{split}
\alpha(F_{\mathds 1}) = 1 \, , \qquad 
\alpha(F_{\II}) \ge 0 \, , \qquad 
P^{[0,0]} \left( \Delta_{[0,0]}' + x \right) \ge 0 \, , \qquad 
P^{[1,0]} \left( \Delta_{[1,0]}' + x \right) \ge 0 \, ,
\end{split}
\end{align}
for all $x \ge 0$.
This is a semidefinite programming problem which can be solved using \texttt{SDPB}~\cite{Simmons-Duffin:2015qma}.

\begin{figure}
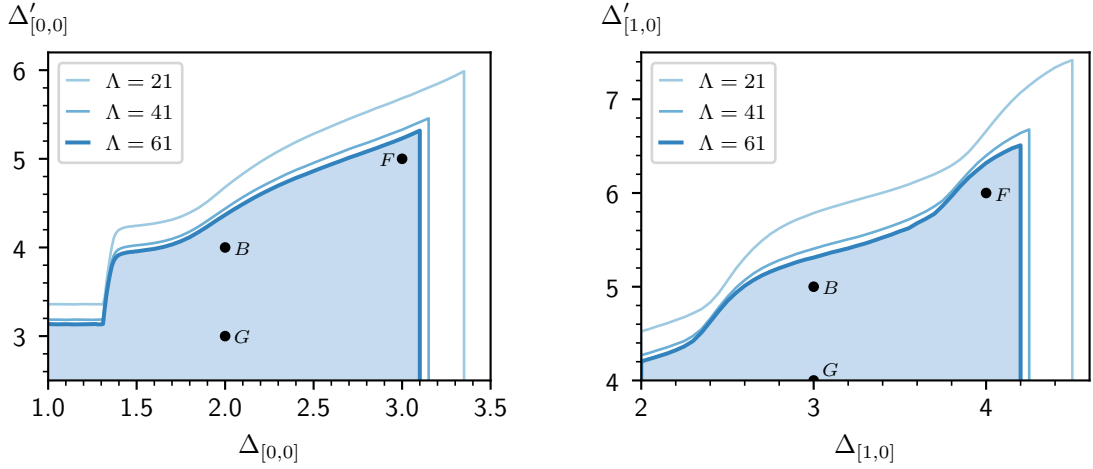

\centering
\begin{subfigure}{.5\textwidth}
  \begin{center} 
   \input{./figures/BoundDelta00pVsDelta00.pgf}
  \end{center}
\end{subfigure}%
\begin{subfigure}{.5\textwidth}
  \begin{center}
   \input{./figures/BoundDelta10pVsDelta10.pgf}
  \end{center} 
\end{subfigure} 
\caption{
Left: Upper bounds on the dimension $\Delta_{[0,0]}'$ of the first long in the continuum as a function of the dimension $\Delta_{[0,0]}$ of the isolated long. 
Only the allowed region for $\Lambda = 61$ is shaded. 
There is a sudden jump in the upper bound for $\Delta_{[0,0]} \simeq 1.31$. 
We are not imposing any gaps in the channel $[1,0]$, and we keep the operators slightly above the unitarity bound, i.e. $\Delta_{[1,0]} = \Delta_{[1,0]}' \gtrsim 2$.
Right: Upper bounds on $\Delta_{[1,0]}'$ as a function of the dimension $\Delta_{[1,0]}$ keeping $\Delta_{[0,0]} = \Delta_{[0,0]}' \gtrsim 1$.
The free theory solutions are represented by bullets $\bullet$, as explained in~\eqref{eq:free-solutions-bullets}.}
\label{fig:bound-dimensions}
\end{figure}

In figure~\ref{fig:bound-dimensions} we present upper bounds on the dimension $\Delta'_{[0,0]}$ of the first long in the continuum, as a function of the dimension of the isolated long $\Delta_{[0,0]}$, while keeping all the operators in the $[1,0]$ channel slightly above their unitarity bound.
In an exactly analogous way, we also present the upper bound of $\Delta'_{[1,0]}$ as a function of $\Delta_{[1,0]}$ without imposing gaps in the $[0,0]$ channel.
The first interesting feature is that regardless of where the continuum sits, there is an upper bound on the dimension $\Delta_{[a,b]}$ of the first long.
The plots suggest that in the limit $\Lambda \to \infty$ the maximum dimension is approximately\footnote{%
It would be interesting to confirm that for larger values of $\Lambda$ the bounds indeed converge to 
$\Delta_{[0,0]} = 3$ and $\Delta_{[1,0]} = 4$, but at this stage the assumption is very plausible.
}
\begin{align}
\label{eq:bounds-delta00-delta10}
 \Delta_{[0,0]} \lesssim 3.0 \, , \qquad
 \Delta_{[1,0]} \lesssim 4.0 \, .
\end{align}
These bounds are almost saturated by the fermionic free theory of equation~\eqref{eq:free-field-theory} with $\xi = -1$.
Moreover, the fermionic theory sits very close to the upper bound for $\Delta_{[0,0]}'$ and $\Delta_{[1,0]}'$ when~\eqref{eq:bounds-delta00-delta10} is saturated.
Similarly, we also see that when $\Delta_{[0,0]} = 2.0$ or $\Delta_{[1,0]} = 3.0$, the free bosonic theory almost saturates the upper bounds for $\Delta_{[0,0]}'$ and $\Delta_{[1,0]}'$ respectively.
Finally, the free gauge theories~\eqref{eq:free-field-theory} with $-1 < \xi < 1$ are far from the boundary of the allowed region.

Another feature is the sudden jump in the upper bound for  $\Delta'_{[0,0]}$ starting at
\begin{align}
\label{eq:delta-jump}
 \Delta_{[0,0],\text{jump}} \simeq 1.31 \, .
\end{align}
As we will discuss in more detail in the following section, this seems to be related to certain OPE coefficients becoming unbounded for $\Delta_{[0,0]} < \Delta_{[0,0],\text{jump}}$.

\subsection{OPE bounds}

\begin{figure}
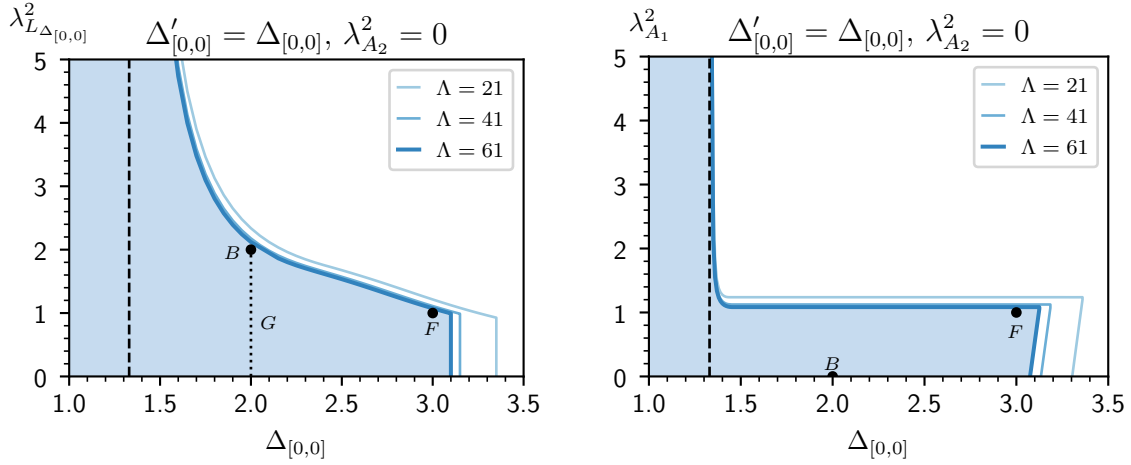

\centering
\begin{subfigure}{.5\textwidth}
  \begin{center} 
   \input{./figures/BoundLambdaDDLVsDelta00WithoutA2.pgf}
  \end{center} 
\end{subfigure}%
\begin{subfigure}{.5\textwidth}
  \begin{center}
   \input{./figures/BoundLambdaDDA1VsDelta00WithoutA2.pgf}
  \end{center}
\end{subfigure}
\caption{
Left: Upper bound on the OPE coefficient of the isolated long as a function of its dimension $\Delta_{[0,0]}$.
Right: Upper bound on the OPE coefficient of the $[A_1]^{j=1/2}_{R=1/2}$ multiplet, as a function of the dimension of the first long $\Delta_{[0,0]}$.
In both plots, we keep $\Delta_{[0,0]}' \gtrsim \Delta_{[0,0]}$ and $\Delta_{[1,0]} = \Delta_{[1,0]}' \gtrsim 2$.
The upper bound of both OPE coefficients diverges for $\Delta_{[0,0]} \simeq 1.33$, which is represented with a vertical dashed line.} 
\label{fig:bound-lambda-vs-delta}
\end{figure}

\begin{figure}
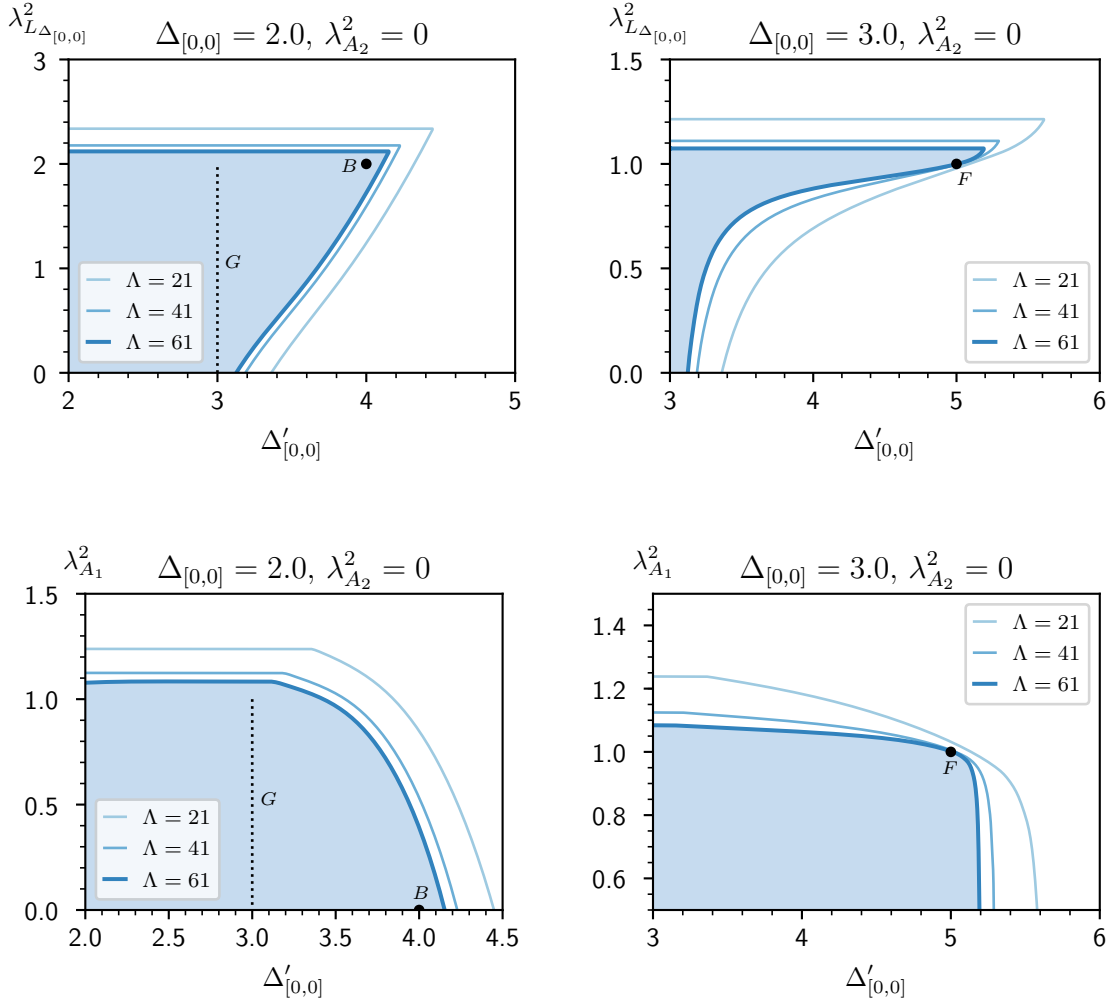

\centering
\begin{subfigure}{.5\textwidth}
  \begin{center} 
   \input{./figures/BoundLambdaDDLVsDelta00pWithoutA2BosonFreeTheory.pgf}
  \end{center}
\end{subfigure}%
\begin{subfigure}{.5\textwidth}
  \begin{center}
   \input{./figures/BoundLambdaDDLVsDelta00pWithoutA2FermionFreeTheory.pgf}
  \end{center}
\end{subfigure}
\begin{subfigure}{.5\textwidth}
  \begin{center} 
   \input{./figures/BoundLambdaDDA1VsDelta00pWithoutA2BosonFreeTheory.pgf}
  \end{center}
\end{subfigure}%
\begin{subfigure}{.5\textwidth}
  \begin{center}
   \input{./figures/BoundLambdaDDA1VsDelta00pWithoutA2FermionFreeTheory.pgf}
  \end{center}
\end{subfigure}
\caption{Upper and lower bounds for $\lambda^2_{L^{[0,0]}_\Delta}$ (first row) and $\lambda^2_{A_1}$ (second row) as a function of $\Delta_{[0,0]}'$ when $\lambda^2_{A_2} = 0$.
In the first column, $\Delta_{[0,0]} = 2.0$ and by increasing $\Delta_{[0,0]}'$ the bosonic free theory sits at the boundary of the allowed region.
In the second column, $\Delta_{[0,0]} = 3.0$ and by increasing $\Delta_{[0,0]}'$ the fermionic free theory sits at the boundary.
} 
\label{fig:bound-lambda-vs-deltaP}
\end{figure}

\begin{figure}
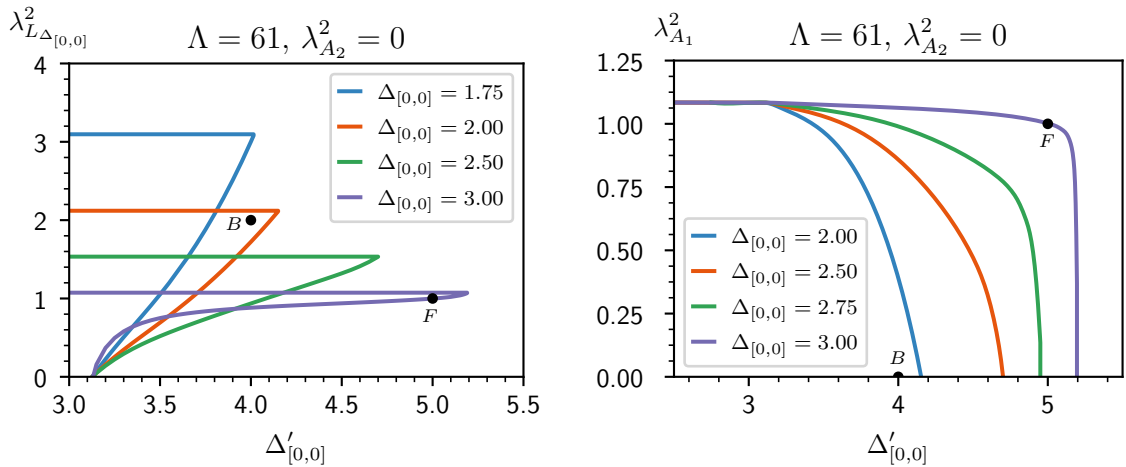

\centering
\begin{subfigure}{.5\textwidth}
  \begin{center} 
   \input{./figures/BoundLambdaDDLVsDelta00pWithoutA2CompareDelta00.pgf}
  \end{center}
\end{subfigure}%
\begin{subfigure}{.5\textwidth}
  \begin{center}
   \input{./figures/BoundLambdaDDA1VsDelta00pWithoutA2CompareDelta00.pgf}
  \end{center}
\end{subfigure}
\caption{
Comparison of the upper and lower bounds of $\lambda^2_L$ (left) and $\lambda^2_{A_1}$ (right) as a function of $\Delta_{[0,0]}'$ and for different values of $\Delta_{[0,0]}$.
All the optimizations have been run for $\Lambda = 61$ and assuming $\lambda_{A_2}^2 = 0$.} 
\label{fig:bound-lambda-vs-deltaP-diffDelta}
\end{figure}

One can find upper and lower bounds for the OPE coefficient $\lambda^2_{\OO}$ using a very similar algorithm as the one described above.
We use a functional $\alpha$ of the form~\eqref{eq:approx-conf-blocks}, and maximize $\alpha(F_{\mathds 1})$ such that $\alpha(F_\OO) = 1$ and
\begin{align}
\label{eq:alg-bound-ope-coef}
 \alpha(F_\II) \ge 0, \qquad
 P^{[0,0]} \left( \Delta_{[0,0]}' + x \right) \ge 0 \;\; \text{for} \;\; x \ge 0 \, , \qquad 
 P^{[1,0]} \left( \Delta_{[1,0]}' + x \right) \ge 0 \;\; \text{for} \;\; x \ge 0 \, .
\end{align}
Then we obtain the upper bound $\lambda^2_{\OO} \le -\alpha(F_{\mathds{1}})$.
Similarly, if we find $\alpha$ that maximizes $\alpha(F_{\mathds 1})$ such that $\alpha(F_\OO) = -1$ and~\eqref{eq:alg-bound-ope-coef} holds, we obtain the lower bound $\lambda^2_{\OO} \ge \alpha(F_{\mathds{1}})$.
As before, such optimization problems can be solved using \texttt{SDPB}.

First, we would like to understand the nature of the jump observed in figure~\ref{fig:bound-dimensions} and discussed around equation~\eqref{eq:delta-jump}.
In figure~\ref{fig:bound-lambda-vs-delta} we obtain upper and lower bounds on the OPE coefficients $\lambda_{L_{\Delta_{[0,0]}}}^2$ and $\lambda_{A_1}^2$ as a function of the dimension of the first long $\Delta_{[0,0]}$.
Here, we are not assuming a double gap in any of the two long channels, i.e. we take $\Delta_{[a,b]} = \Delta_{[a,b]}'$, but we do assume $\lambda_{A_2}^2 = 0$.
Somehow unexpectedly, both OPE coefficients become unbounded for $\Delta_{[0,0]}$ less than
\begin{align}
\label{eq:delta-jump-2}
  \Delta_{[0,0],\text{jump}} \simeq 1.33 \, .
\end{align}
Even though there is a slight mismatch between the values of $\Delta_{[0,0],\text{jump}}$ in~\eqref{eq:delta-jump} and~\eqref{eq:delta-jump-2}, we believe it is only due to the numerical nature of the calculation, and that the two values would be the same for large enough $\Lambda$.
A very similar situation was observed in~\cite{Liendo:2018ukf}, where a sudden drop in the upper bound of a conformal dimension was related to the appearence of an upper bound of a related OPE coefficient.
For the $3d$ Ising model it is known that the dimensions and OPE coefficients of certain operators suffered a sudden jump around the Ising model point~\cite{El-Showk:2014dwa}.
It would be interesting to see if the region $\Delta_{[0,0]} \sim \Delta_{[0,0],\text{jump}}$ corresponds to a line defect of an interesting $\NN = 2$ superconformal theory.

In order to obtain further constraints on OPE coefficients we will assume the existence of gaps, in particular, $\Delta_{[0,0]} \ge \Delta_{[0,0],\text{jump}}$, because otherwise the optimization problems are unbounded.
As an important example, we study in more detail the exact bosonic and fermionic solutions of crossing.
We fix the dimension of the first long to $\Delta_{[0,0]} = 2.0/3.0$ for the bosonic/fermionic theories, and then bound the OPE coefficients as we increase the second gap $\Delta_{[0,0]}'$.
The results are plotted in figure~\ref{fig:bound-lambda-vs-deltaP}.
In the first row we observe that the OPE coefficient of the long at $\Delta_{[0,0]}$ has upper bounds which are essentially constant, and lower bounds appear only when the second gap is $\Delta_{[0,0]}' \gtrsim 3$.
The lower bounds grow as we increase $\Delta'_{[0,0]}$, until they meet the upper bound precisely where the bosonic and fermionic theories sit.
For this reason, we expect that the bosonic and fermionic theories are unique provided that the second gap is large enough.
Indeed, our plots are almost identical to the ones obtained for the $\NN = 4$ analogous case~\cite{Liendo:2018ukf}.
In order to map results, one simply needs to note that their $\BB_2$ multiplet is identified with our isolated long of dimension $\Delta_{[0,0]} = 2$ (see the discussion around equation~\eqref{eq:decompose-N4-N2}).
A mixed-correlator bootstrap study for $\NN = 4$ revealed the appearence of an island around the bosonic free theory.
We are confident that a similar analysis can be done in our setup, which would give evidence that our free-field solutions of crossing are unique if one assumes appropriate gaps.

In the second row of figure~\ref{fig:bound-lambda-vs-deltaP} we show bounds on the OPE coefficient of the $[A_1]_{R=1/2}^{j=1/2}$ multiplet.
There is no analogous of this multiplet for line defects in $\NN = 4$ theories, so we will not be able to borrow any intuition from the results of~\cite{Liendo:2018ukf}. 
The primary of $A_1$ has dimension $\Delta = 5/2$, so it sits inside the continuum of $[1,0]$ long operators.
Intuitively, in order for lower bounds to appear, there needs to be enough distance between the dimension of the operator and the dimension of the first operator in the continuum, and that explains why we do not obtain any lower bounds for $\lambda_{A_1}^2$.
In any case, when $\Delta_{[0,0]} = 2$ the upper bound keeps decreasing until it crosses zero, exactly at the position where the bosonic free theory sits.
When $\Delta_{[0,0]} = 3$, the bounds seems to converge to the rectangular region $\lambda_{A_1}^2 \le 1$ and $\Delta_{[0,0]}' \le 5$, and the fermionic theory sits exactly at the upper right corner of this region. 

Summarizing, figure~\ref{fig:bound-lambda-vs-deltaP} provides ample evidence that the numerical bootstrap is isolating the bosonic and fermionic free theories when we assume large gaps in the spectrum of long operators.
Interestingly, one can allow the dimension for the first long to be in the range
\begin{align}
\Delta_{\text{jump},[0,0]} \le \Delta_{[0,0]} \le 3,
\end{align}
and compute bounds on OPE coefficients as a function of $\Delta_{[0,0]}'$.
The results are plotted in figure~\ref{fig:bound-lambda-vs-deltaP-diffDelta}.
There is an entire family of plots that share similar qualitative features to the ones we just discussed.
This can be thought of as an one-parameter family of theories interpolating between the fermionic and bosonic free-field theories, and which would extend all the way up to the critical theory where the OPE coefficients are diverging.

%!TEX root = ../N2_line.tex
%%%%%%%%%%%%%%%%%%%%%%%%%%%%%%%%%%%%%%%%%%%%%
\section{Analytical results}
\label{sec:analytical}

\subsection{Introduction}

\begin{figure}
  \centering
  \def\svgwidth{0.6\textwidth}
%% Accompanies image file 'witten_diagrams.pdf' (pdf, eps, ps)
\begingroup%
  \makeatletter%
  \providecommand\color[2][]{%
    \errmessage{(Inkscape) Color is used for the text in Inkscape, but the package 'color.sty' is not loaded}%
    \renewcommand\color[2][]{}%
  }%
  \providecommand\transparent[1]{%
    \errmessage{(Inkscape) Transparency is used (non-zero) for the text in Inkscape, but the package 'transparent.sty' is not loaded}%
    \renewcommand\transparent[1]{}%
  }%
  \providecommand\rotatebox[2]{#2}%
  \newcommand*\fsize{\dimexpr\f@size pt\relax}%
  \newcommand*\lineheight[1]{\fontsize{\fsize}{#1\fsize}\selectfont}%
  \ifx\svgwidth\undefined%
    \setlength{\unitlength}{481.98816495bp}%
    \ifx\svgscale\undefined%
      \relax%
    \else%
      \setlength{\unitlength}{\unitlength * \real{\svgscale}}%
    \fi%
  \else%
    \setlength{\unitlength}{\svgwidth}%
  \fi%
  \global\let\svgwidth\undefined%
  \global\let\svgscale\undefined%
  \makeatother%
  \begin{picture}(1,0.52728042)%
    \lineheight{1}%
    \setlength\tabcolsep{0pt}%
    \put(0,0){\includegraphics[width=\unitlength,page=1]{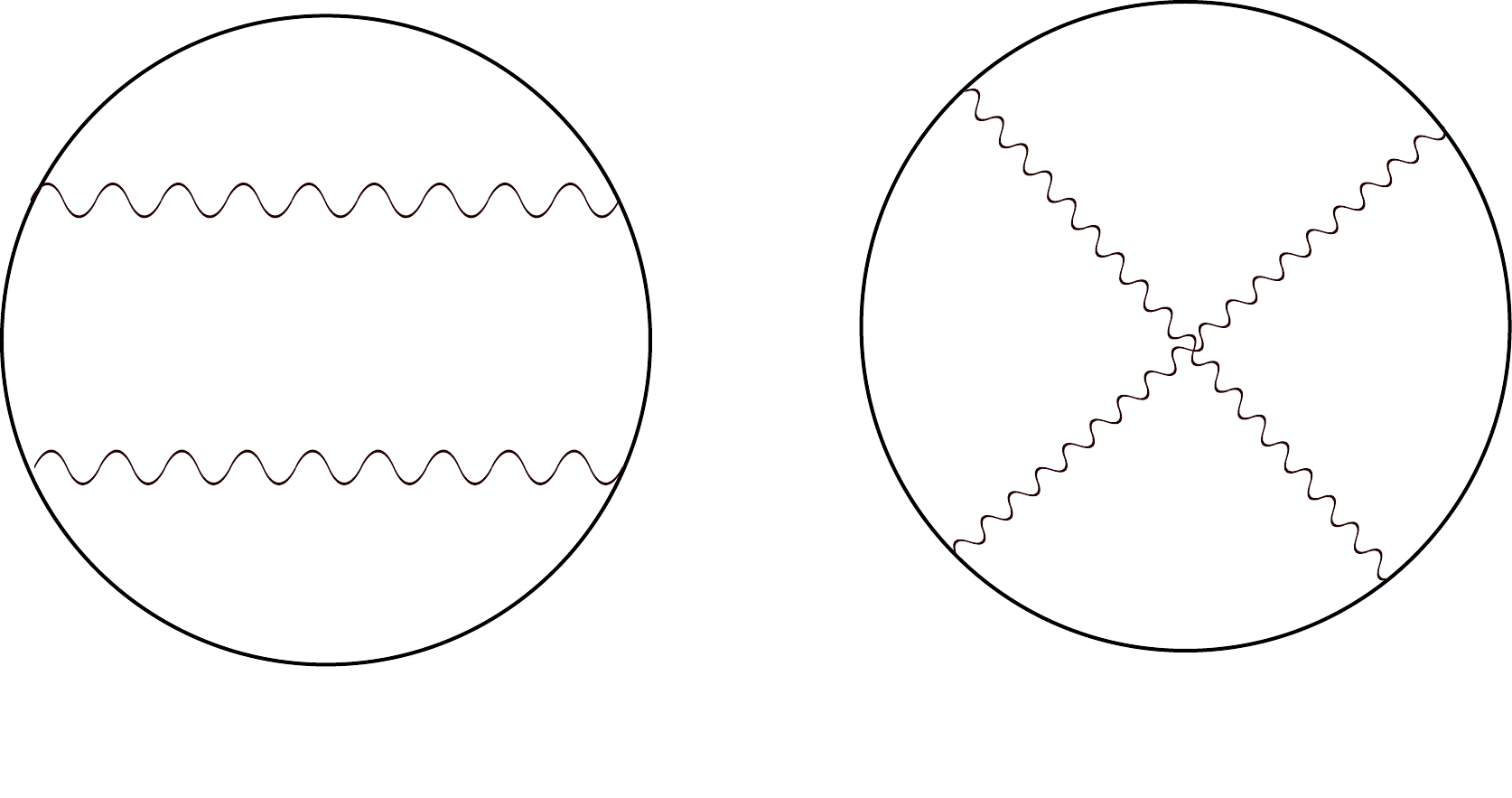}}%
    \put(0.18588157,0.00820575){\color[rgb]{0,0,0}\makebox(0,0)[lt]{\lineheight{1.25}\smash{\begin{tabular}[t]{l}\textbf{a)}\end{tabular}}}}%
    \put(0.76589505,0.0113494){\color[rgb]{0,0,0}\makebox(0,0)[lt]{\lineheight{1.25}\smash{\begin{tabular}[t]{l}\textbf{b)}\end{tabular}}}}%
  \end{picture}%
\endgroup%

  \caption{Disconnected and connected Witten diagrams in the dual $AdS_2$ description. The disconnected piece corresponds to a mean-field theory correlator, while the connected piece is bootstrapped in the current section.}
  \label{fig:witten}
\end{figure}

In this section we study perturbations around the bosonic mean-field solution~\eqref{eq:free-field-theory},  similar to the analysis of section 6 in~\cite{Liendo:2018ukf}. We will interpret the bosonic solution as the strong-coupling limit of line defects in $\Nm=2$ theories which admit a holographic description. 
From the holographic perspective, the leading contribution to a four-point function at strong coupling is a disconnected Witten diagram in $AdS_2$, while the first-order correction is given by a four-point connected Witten diagram, see figure~\ref{fig:witten}.
The disconnected piece can be obtained by Wick contractions, and leads to our solution~\eqref{eq:free-field-theory} with $\xi = 1$.
Our goal is to use superconformal blocks and crossing symmetry to bootstrap the contribution from the connected Witten diagram.
We will see that under mild assumptions, the correlator is uniquely determined in terms of two normalization constants $c_1, c_2$, which cannot be fixed by our symmetry arguments.
From the correlator it is then possible to extract the first-order corrections to the anomalous dimensions and OPE coefficients of the operators in the spectrum.
In the analogous $\Nm=4$ case, perfect agreement was found between the explicit holographic calculation~\cite{Giombi:2017cqn} and the bootstrap result~\cite{Liendo:2018ukf}.

Let us remind the reader that in section~\ref{sec:crossing-equations} we wrote the crossing equation~\eqref{eq:crossing-equation} in terms of the two-dimensional vector $H(z)$.
This function can be expressed in a superblock-like expansion
\beq
\label{eq:block_expansion}
H(z) = \sum_{\Delta \in S^{[0,0]}}a_{\Delta} H^{[0,0]}_{\Delta}(z) 
+ \sum_{\Delta \in S^{[1,0]}}b_{\Delta} H^{[1,0]}_{\Delta}(z),
\eeq
where $H^{[a,b]}_{\Delta}$ are also two-dimensional vectors that can be computed from the definition of $H(z)$ in~\eqref{eq:define-H} and the superconformal blocks in the two channels~\eqref{eq:sol_SSSS_00} and~\eqref{eq:sol_SSSS_10}.
One can think of $H^{[a,b]}_{\Delta}$ as a superblock expressed in a new basis, such that the crossing equation takes a particularly simple form.

The solution to crossing we want to perturb around has OPE coefficients given in equation~\eqref{eq:ope-coefs-free-theory} with $\xi=1$, and the spectrum of dimensions is
\begin{align}
 S_{[0,0]} = \{2,4,6, \dots\}  
 \qquad \text{and} \qquad 
 S_{[1,0]} = \{3,5,7, \dots\}.
\end{align}
The idea is to start with this free theory and consider a perturbation of the CFT data to leading order in the perturbation parameter $\epsilon$.
On the one hand, the correlator will receive a correction
\begin{align}
 H(z) & = H^{(0)}(z) + \veps H^{(1)}(z)\, , 
\end{align}
which by equation~\eqref{eq:block_expansion} will translate into the operators acquiring anomalous dimensions
\begin{align}
 S^{(1)}_{[0,0]} 
 = \{\Delta + \veps \gamma^{[0,0]}_{\Delta} \}_{\Delta \in S_{[0,0]}}, \qquad
 S^{(1)}_{[1,0]} 
 = \{\Delta + \veps \gamma^{[1,0]}_{\Delta} \}_{\Delta \in S_{[1,0]}},
\end{align}
and the OPE coefficients receiving first-order corrections
\begin{align}
a_{\Delta} = a^{(0)}_{\Delta} + \veps a^{(1)}_{\Delta}, \qquad
b_{\Delta} = b^{(0)}_{\Delta} + \veps b^{(1)}_{\Delta}.
\end{align}
Schematically, we have that $H^{[a,b]}_\Delta \sim z^\Delta f(\Delta,z)$, so if we give an anomalous dimension to $\Delta$ the first-order correlator $H^{(1)}(z)$ must contain a log term.
As a result, we take it to be of the form
\begin{align}
 \label{eq:ansatz-R}
 H^{(1)}(z) =  R(z) \log(z) + P(z)\,,
\end{align}
where $R(z)$ and $P(z)$ are a priory completly arbitrary functions.
Comparing this with the block expansion we obtain
\begin{subequations}
\label{eq:expansion-RP}
 \begin{align} 
\label{eq:expansion-R}
R(z) & = \sum_{\Delta \in S_{[0,0]}}a^{(0)}_{\Delta} \gamma^{[0,0]}_{\Delta}H^{[0,0]}_{\Delta}(z) 
+ \sum_{\Delta \in S_{[1,0]}}b^{(0)}_{\Delta} \gamma^{[1,0]}_{\Delta} H^{[1,0]}_{\Delta}(z),
\\
\begin{split}
\label{eq:expansion-P}
P(z) 
& = \sum_{\Delta \in S_{[0,0]}}a^{(1)}_{\Delta} H^{[0,0]}_{\Delta}(z) 
  + \sum_{\Delta \in S_{[0,0]}}a^{(0)}_{\Delta} 
    \gamma^{[0,0]}_{\Delta}  
    z^{\Delta} \pd_\Delta \left( z^{-\Delta} H^{[0,0]}(z) \right) \\
& + \sum_{\Delta \in S_{[1,0]}}b^{(1)}_{\Delta} H^{[1,0]}_{\Delta}(z)
  + \sum_{\Delta \in S_{[1,0]}}b^{(0)}_{\Delta} 
    \gamma^{[1,0]}_{\Delta}
    z^{\Delta} \pd_\Delta \left( z^{-\Delta} H^{[1,0]}(z) \right),
\end{split}
\end{align}
\end{subequations}

In the analysis below, the ``brading'' transformation
\beq
\label{eq:braiding} 
z \to \frac{z}{z-1}
\eeq
will play a crucial role to provide extra constraints for the functions $R(z)$ and $P(z)$.
The one-dimensional bosonic blocks $g_\Delta = g_\Delta^{0,0}$ of equation~\eqref{eq:chiral-block} have clean transformation properties under braiding.
In our analysis, only chiral blocks with even $\Delta$ will appear, for which we have\footnote{%
For generic values of $\Delta$, the chiral block will have an extra branch cut due to the prefactor $z^\Delta$, and one has to be careful on how to analytically continue the block under~\eqref{eq:braiding}.
See \cite{Bissi:2018mcq} for a careful analysis in the BCFT setup.
}
\begin{align}
\label{eq:braiding-chiral-block}
 g_\Delta \left( \frac{z}{z-1} \right) = g_\Delta(z), \qquad
 g_\Delta' \left( \frac{z}{z-1} \right)
 = - (1-z)^2 g_\Delta'(z), \qquad
 \text{etc.}
\end{align}
From the form of the superconformal blocks $G_i(z)$, it is clear that they inherit these nice transformation properties under braiding.
However, when we work in the $H$-basis, the transformations become more complicated
and instead of writing them here we will only present their consequences. 
Using the transformation~\eqref{eq:braiding-chiral-block} combined with the expansions~\eqref{eq:expansion-RP}, we obtain non-trivial constraints for the two components of  $R(z)$ 
\begin{align}
\label{eq:braiding-R}
 R_0 \left( \frac{z}{z-1} \right) - R_0(z) = 0, \qquad
 R_1 \left( \frac{z}{z-1} \right) - \BB[R](z) = 0,
\end{align}
and for the two components of $P(z)$
\begin{subequations}
\label{eq:braiding-P}
\begin{align}
 \label{eq:braiding-P0}
 & P_0 \left( \frac{z}{z-1} \right) - P_0(z) - \log(1-z) R_0(z) = 0, \\
 \label{eq:braiding-P1}
 & P_1 \left( \frac{z}{z-1} \right) - \BB[P](z) - \log(1-z) \BB[R](z) + \frac{z}{z-1} R_0(z) = 0.
\end{align}
\end{subequations}
Here we have defined a functional $\BB$, which takes as argument a two-component function $F(z)$ and mixes its two components as follows:
\begin{align}
 \BB[F](z) = 
 - \frac{2 z (z - 2) }{(z-1)^2} F_0(z)
 - \frac{z^2}{z-1} \pd_z F_0(z)
 + \frac{1}{(z-1)^2} F_1(z)\,.
\end{align}
In the next section we will study how these constraints fix the functions $R(z)$ and $P(z)$ up to overall coefficients.

\subsection{Corrections to the anomalous dimension}

We are now ready to find solutions to crossing which are consistent with the relations just presented.
In order to do so, we take the function $P(z)$ to be of the form
\begin{align}
 \label{eq:ansatz-P}
 P(z) = \frac{z^2}{(1-z)^2} R(1-z) \log(1-z) + Q(z),
\end{align}
and we assume that $R(z)$ and $Q(z)$ are rational functions.
This assumption is inspired by the holographic calculation of \cite{Giombi:2017cqn}, and can also be justified a posteriori if a solution is actually found.
The idea is that the contribution from the connected Witten diagram in figure~\ref{fig:witten} is given by an integral of four bulk-to-boundary propagators living in $AdS_2$, which is denoted by $D_{\Delta_1\Delta_2\Delta_3\Delta_4}$ in~\cite{Giombi:2017cqn}.
For the case of interest to us, the conformal dimensions of the external operators are all identical and take integer values, in which case the only transcendental functions appearing in $D$ are $\log(z)$ and $\log(1-z)$.
Therefore, our ansatz~\eqref{eq:ansatz-R} and~\eqref{eq:ansatz-P} is the most general one representing a first order correction in the holographic dual.

Due to the form of our ansatz, crossing symmetry does not impose conditions on the function $R(z)$, however the braiding property does impose non-trivial relations on both $R(z)$ and $Q(z)$. 
It turns out that it is sufficient to solve~\eqref{eq:braiding-P} and that~\eqref{eq:braiding-R} does not impose extra constraints.
Also, recall that under the assumption of rationality the coefficients of possible log terms have to cancel separately.
Now we insert our ansatz~\eqref{eq:ansatz-P} in \eqref{eq:braiding-P0}, and by extracting the coefficient of the log term, we obtain the following relation for the function $R_0(z)$:
\beq
-z^2 R_0\left(\frac{1}{1-z}\right)-\frac{z^2 R_0(1-z)}{(z-1)^2}-R_0(z) = 0.
\eeq
Similarly, by looking at~\eqref{eq:braiding-P1} we obtain an equation that mixes the two components of $R(z)$
\begin{align}
\begin{split}
-(z-1) z^4 R_0'(1-z)+2 (z-3) z^3 R_0(1-z)+(z-1)^3 z^2 R_0'(z)+2 (z-2) (z-1)^2 z R_0(z)
\\
-z^2 (z-1)^4 R_1\left(\frac{1}{1-z}\right)-z^2 R_1(1-z)+\left(-z^2+2 z-1\right) R_1(z) & = 0.
\end{split}
\end{align}
In addition to these relations, the function $R(z)$ is constrained by the block expansion~\eqref{eq:expansion-RP}. In particular, in the limit $z \sim 0$ it should satisfy 
\beq
(R_0(z),R_1(z)) \sim (z^2,-2 z^2),
\eeq
where the relative factor of $-2$ comes from the explicit normalization of the conformal blocks in the basis we employ. 
As discussed in \cite{Liendo:2018ukf}, these conditions are not enough to fix the function $R(z)$, and we need to look at the behavior of the function around $z \sim 1$, which is correlated with the behavior of anomalous dimensions at large $\Delta$. Because we are looking for a solution that can be interpreted as a holographic correlator, we will borrow some intuition from \cite{Heemskerk:2009pn,Fitzpatrick:2010zm}. The idea is that the growth of anomalous dimensions is governed by how irrelevant the interaction is in the putative $AdS$ dual. Because we are trying to bootstrap a \textit{leading} correction to the holographic correlator, we should keep the solution with the \textit{weakest} growth. Therefore, we impose that anomalous dimensions grow no faster than $\gamma_{\Delta}^{[a,b]} \sim \Delta^2$ for large values of $\Delta$. This last condition fixes the function $R(z)$ up to two normalization constants. The explicit answer reads
\begin{align}
R_0(z) & = -\frac{z^2}{z-1} c_1 -\frac{\left(2 z^2-7 z+7\right) z^4}{2 (z-1)^3} c_2\, ,
\\
R_1(z) & = -\frac{z^2 \left(2 z^2-3 z-6\right)}{3 (z-1)}c_1 + \frac{z^4 \left(8 z^2-28 z+35\right)}{3 (z-1)^3} c_2\, .
\end{align}
It is instructive to compare this result with the analysis of \cite{Liendo:2018ukf} for line defects in $\Nm=4$ theories. In the $\Nm=4$ case, there is only one function and the solution could be fixed up to an overall coefficient. Moreover, this coefficient is associated to a three-point function of half-BPS operators and can be fixed using localization \cite{Giombi:2018qox}. In our case of line defects in $\Nm=2$ theories, we have two overall constants associated to each independent channel. Unlike $\Nm=4$ SYM, which seems to be unique, we know that there is an extensive catalog of $\Nm=2$ theories, and it is then no surprise that our solution has more freedom.

From the explicit solution for $R(z)$, the anomalous dimensions can be read off from the block expansion in \eqref{eq:expansion-R}:
\begin{align}
\gamma_{\Delta}^{[0,0]} & =\frac{\Delta  (\Delta +1)}{3 (\Delta -1) (\Delta +2)}c_1 +\frac{(\Delta -2) (\Delta +3) \big(3 \Delta(\Delta+1) -4 \big)}{12 (\Delta -1) (\Delta +2)}c_2\, ,
\\
\gamma_{\Delta}^{[1,0]}  & = -\frac{(\Delta -1) (\Delta +2)}{9 \Delta  (\Delta +1)}c_1
+\frac{(\Delta -1) (\Delta +2) \big(9 \Delta(\Delta+1) +4 \big)}{36 \Delta  (\Delta +1)}c_2\, .
\end{align}
From this expression is clear that they scale as $\Delta^2$ for large $\Delta$.

\subsection{Corrections to the OPE coefficients}

With the explicit solution for $R(z)$ at hand, we can proceed to compute $Q(z)$, which will allow us to extract the first-order correction to the OPE coefficients.
The crossing equation gives non-trivial constraints for both components of $Q(z)$, namely
\beq  
Q(z)-\frac{z^2}{(1-z)^2}Q(1-z) = 0.
\eeq
The equations coming from braiding will provide extra conditions, in particular if we insert the ansatz~\eqref{eq:ansatz-P} in~\eqref{eq:braiding-P1} and now extract the term with no logs, we get
\begin{align}
\begin{split}
(z-1)^4 Q_1\left(\frac{z}{z-1}\right)-2 (2-z) z (z-1)^2 Q_0(z)-(z-1)^2 Q_1(z)
\\
+z (z-1)^3 R_0(z)+z^2 \left((z-1)^3 Q_0'(z)+z^2 R_0(1-z)\right) & = 0.
\end{split}
\end{align}
As before, the other braiding equations do not provide extra conditions.
It only remains to impose the boundary conditions for $z \sim 0$ similarly to what we did for $R(z)$. Our final solution for  $Q(z)$ is given by 
\begin{align}
Q_0(z)  & = \frac{\left(z^2-z+1\right)^2}{(z-1)^2} c_2,
\\
Q_1(z) & = \frac{2  z^2}{3}c_1-\frac{\left(16 z^4-32 z^3+97 z^2-81 z+30\right)}{6 (z-1)^2} c_2.
\end{align}
Having both $R(z)$ and $Q(z)$, we can now use the block expansion~\eqref{eq:expansion-P} to extract corrections to the OPE coefficients, similarly to what we did for the anomalous dimension. It turns out that the corrections $a^{(1)}_{\Delta}$ and $b^{(1)}_{\Delta}$ can be elegantly written in terms of the derivatives of the anomalous dimensions times the zeroth-order values for $a^{(0)}_{\Delta}$ and $b^{(0)}_{\Delta}$:
\beq
a^{(1)}_{\Delta}  = \frac{\pd}{\pd \Delta} (a^{(0)}_{\Delta} \gamma^{[0,0]}_{\Delta})\, , \qquad
b^{(1)}_{\Delta}  = \frac{\pd}{\pd \Delta} (b^{(0)}_{\Delta} \gamma^{[1,0]}_{\Delta})\, .
\eeq
Similar relations were originally observed in \cite{Heemskerk:2009pn,Fitzpatrick:2011dm}. It is not clear to us which of our assumptions implies these relations, but in any case it is reassuring to see that they are satisfied.

Let us finish with some comments. From the start we are assuming that the spectrum of the perturbed solution is the same as the spectrum of the zeroth-order starting point. In principle, there could be degenerate families that are lifted at first order. However, because we are looking at a single correlator, possible degeneracies are invisible at this stage of the calculation. The more correct way to interpret our results is as weighted averages  \cite{Alday:2017xua,Aprile:2017xsp}. In order to resolve possible degeneracies it is necessary to study a mixed correlator system. For example, one could use the correlators involving long multiplets that we present in appendix~\ref{apx:long-blocks}, although perhaps more general correlators are needed in which the external operators carry non-zero quantum numbers under $\su{(2)}_j \times \su{(2)}_R$. We leave this interesting problem for future work.

Let us also point out that this solution to crossing is interesting in its own right. It would be ideal to compare our result with other approaches and explicit holographic calculations in some selected $\Nm=2$ model, as it would allow us to understand the origin of the coefficients $c_1$ and $c_2$. 
Finally, a similar calculation to ours was done in \cite{Mazac:2018ycv} using the exact functional method, where possible deformations of a free theory were bootstrapped by explicitly constructing the exact functionals that give the optimal bound. It would be interesting to adapat the approach of \cite{Mazac:2018ycv} to our crossing constraints \eqref{eq:crossing-equation}.

%!TEX root = ../N2_line.tex
%%%%%%%%%%%%%%%%%%%%%%%%%%%%%%%%%%%%%%%%%%%%%

\section{Conclusions}
\label{sec:conclusions}
In this work we have initiated the bootstrap program for line defects in $\Nm=2$ theories. We studied the $1d$ CFT that lives in a line defect using a collection of bootstrap techniques. Our results are for the most part very general, as they rely on basic symmetry principles and consistency requirements, and are therefore valid for standard Wilson and 't Hooft lines in gauge theories, but also for more exotic constructions like line defects in non-Lagrangian models \cite{Xie:2013lca,Xie:2013vfa,Cordova:2016uwk}.

We concentrated mostly on correlators of the displacemente operator, but one can also consider more general external multiplets and study a mixed correlator bootstrap. Partial progress towards this goal is already presented in appendix \ref{apx:long-blocks}, where conformal blocks for correlators that include scalar long multiplets as external operators are shown. 
The analysis of this paper shows that not only scalar long multiplets, but also multiplets charged under transverse spin, are generated in the OPE of two displacements.
Therefore, it would be interesting to consider crossing involving long operators that sit in non-trivial representations of the bosonic subalgebra.

As a longer term goal, one could include local operators outside the defect. This is particularly interesting when considering that theories with the same local spectrum can support different line defects~\cite{Aharony:2013hda}. Basic kinematics constraints on two-point functions in the presence of an $\Nm=2$ line have not been calculated yet. A project for the not so distant future would be to consider a mixed system between the bulk stress tensor and the displacement operator, generalizing the analysis of \cite{Bianchi:2018zpb} where the coupling between the displacement and the stress tensor was studied.
It would also be interesting to see bootstrap constraints on possible line defects when assuming a given bulk CFT.

Another interesting follow-up would be to perform holographic calculations in some specific $\Nm=2$ model, in order to compare with  our analytic correlator from section \ref{sec:analytical}. There seems to be no calculation of this sort in the $\Nm=2$ literature. In $\Nm=4$ SYM the holographic calculation of \cite{Giombi:2017cqn} and the bootstrap analysis of \cite{Liendo:2018ukf} are in perfect agreement. We are confident that there will be a similar match in the $\Nm=2$ case.

One more possible avenue is to push the analytic analysis to higher orders in the perturbative expansion. This was done in \cite{Liendo:2018ukf}  for $\Nm=4$, but in order to resolve the important issue of degeneracies a bigger collection of correlators has to be considered. In addition, one could also try to adapt the exact functional machinery developed in \cite{Mazac:2016qev,Mazac:2018mdx,Mazac:2018ycv}. The systems studied in this work have interesting simplifying features, i.e. $1d$ CFTs with a high amount of supersymmetry, and perhaps exact solutions to the crossing equations are within reach.

\acknowledgments
We thank I.~Buric, E.~Lauria, Y.~Linke, V.~Schomerus, and B.~van Rees for useful discussions.
This work is supported by the DFG through the Emmy Noether research group ``The Conformal Bootstrap Program'' project number 400570283.

\appendix
\section{Conventions}
\label{apx:conventions}

In this appendix, we define an index-free notation to contract the fermionic coordinates $\theta_\bA^a$ of our superspace.
These objects have one transverse-spin and one \Rsymmetry index, and since both groups are $\su(2)$, we will need to use the totally antisymmetric symbol
\begin{align}
     \ee^{12} 
 = - \ee^{21} 
 = - \ee_{12} 
 =   \ee_{21} = 1, \qquad
     \ee^{\bs 1 \bs 2} 
 = - \ee^{\bs 2 \bs 1} 
 = - \ee_{\bs 1 \bs 2} 
 =   \ee_{\bs 2 \bs 1} = 1.
\end{align}
As usual, the conventions to raise or lower indices are as follows
\begin{align}
  \theta_{\bA,\ia} = \ee_{\ia\ib} \theta_\bA^\ib, \qquad
  \theta^{\bA,\ia} = \ee^{\bA\bB} \theta_\bB^\ia, \qquad
  \text{etc.}
\end{align}
There is only one meaningful way to contract two coordinates and form a scalar
\begin{align}
\label{eq:two-pt-contr}
 \theta \xi \equiv \ee_{\ia\ib} \ee^{\bA \bB} \theta_\bA^\ia \xi_\bB^\ib.
\end{align}
Note that $\theta \xi = - \xi \theta$ and therefore $\theta \theta = 0$.
Given three coordinates $\theta$, $\xi$ and $\zeta$, they can be contracted as
\begin{align}
\label{eq:three-pt-contr}
 (\theta \xi \zeta)_{\bA}^\ia 
 = \ee_{\ib\ic} \ee^{\bB\bC}
   \theta_\bA^\ib \xi_\bB^\ia \zeta_\bC^\ic.
\end{align}
This contraction is interesting because it is inequivalent to contracting two coordinates as in~\eqref{eq:two-pt-contr} and then multiplying by the third one.
As a result, it does not vanish even if two or three coordinates are the same: $(\theta \theta \theta)_\bA^\ia \equiv (\theta^3)_\bA^\ia \ne 0$.
Finally, given four Grassmann variables there is one contraction such that it cannot be decomposed as a product of terms of the form~\eqref{eq:two-pt-contr}
\begin{align}
\label{eq:four-pt-contr}
 \theta \xi \zeta \eta 
 = \ee_{\ia\ic} \ee_{\ib\id} \ee^{\bA\bB} \ee^{\bC\bD}
   \theta_\bA^\ia \xi_\bB^\ib \zeta_\bC^\ic \eta_\bD^\id.
\end{align}
As before, this does not vanish even in the case of four identical coordinates $\theta \theta \theta \theta \equiv \theta^4 \ne 0$.
Note also that we could have defined it as $\theta \xi \zeta \eta \equiv \theta_\bA^\ia (\xi \zeta \eta)^\bA_\ia$.

When we classify all possible fermionic invariants, the following relations will be useful
\begin{align}
\label{eq:four-contract-properties}
\begin{split}
 & \theta \xi \theta \xi = \xi \theta \xi \theta, \\
 & \theta \theta \xi \xi = \xi \xi \theta \theta, \\
 & \theta \xi \xi \theta 
 = \xi \theta \theta \xi
 = \tfrac{1}{2} (\theta \xi \theta \xi + \theta \theta \xi \xi), \\
 & \xi \theta \theta \theta
 = \theta \xi \theta \theta
 = \theta \theta \xi \theta
 = \theta \theta \theta \xi,
\end{split}
\end{align}
and also
\begin{align}
\label{eq:extra-contract-properties}
\begin{split}
 (\theta \xi)^2 
 = \tfrac{1}{2} (\theta \xi \theta \xi - \theta \theta \xi \xi ), \qquad
 (\theta \xi)^3
 = - \tfrac{2}{3} \theta^3 \xi^3.
\end{split}
\end{align}

\section{Long blocks}
\label{apx:long-blocks}

In this appendix, we compute superconformal blocks involving unprotected operators.
We start by obtaining the blocks of two displacements and two longs in the $(12)\to(34)$ channel, and then proceed to compute the same blocks involving four long operators.
In order to study crossing for the full mixed system, 
one would still need to compute the blocks $\langle \DD \DD \OO \OO \rangle$ in the $(14)\to(23)$ channel, but we expect this not to be hard using the techniques presented in the paper.

\subsection{Two displacements and two longs}

We will start by computing the superconformal blocks of two displacements $\DD(z)$ with two identical long scalar operators $\OO(z)$ of dimension $\Delta_\OO$ in the $(12)\to(34)$ channel
\begin{align}
\label{eq:expansion-superconf-blocks-SSLL}
 \langle \dispM(z_1) \dispM(z_2) \longM(z_3) \longM(z_4) \rangle 
 = \frac{1}{Z_{12}^2 Z_{34}^{2\Delta_\OO}} 
   \sum_{\OO'} \lambda_{\DD\DD\OO'} \lambda_{\OO\OO\OO'} \, \GG_{\OO'}(I_a).
\end{align}
The steps of the calculation are analogous to section~\ref{sec:superblocks-casimir}, with the exception that now the shortening conditions are given by~\eqref{eq:shorteningSSLL} only.
Therefore, there are three free functions $f_0(z)$, $f_1(z)$ and $f_9(z)$, and there must be an extra independent Casimir equation.
As before, we apply the Casimir operator $\CC^2_{12}$ to the four-point function in the frame \fr 1 to simplify the computations.
We get one of the original Casimir equations~\eqref{eq:casimirEq_SSSS_a}, together with two new constraints:
\begin{subequations}
\begin{align}
\label{eq:casimirEq_SSLL}
& -z^2 \big[ (z-1) f_0''(z) + f_0'(z) \big]-4 z f_1(z) 
  = \cc \, f_0(z), \\[0.5em]
\begin{split}
&       + 2304 z^3 f_9(z)
        -16 (2 \cc+3 z-10) f_1(z)
        +48 (2-3 z) z f_1'(z)
        -48 (z-1) z^2 f_1''(z) \\
& \quad +8 \big[ (z-1) z+6 \big] f_0'(z)
        +2 \big[ z (5 z-4)-8 \big] z f_0''(z) \\
& \quad -2 (z-1) (z+4) z^2 f_0^{(3)}(z)
        -(z-1)^2 z^3 f_0^{(4)}(z) = 0,
\end{split} \\[0.5em]
\begin{split}
&         (\cc -2)^2 \cc f_0(z)
        + z^2 \big[ 3 \cc^2+2 \cc (6 z-5) + 4 z (9 z-8) \big] f_0'(z) \\
& \quad + z^2 \big[ 3 \cc^2 (z-1)+2 \cc (z (21 z-23)+5)+4 z (7 z-6) (9 z-4) \big] f_0''(z) \\
& \quad + 2 z^3 \big[ 6 \cc (z-1) (2 z-1) + z (z (165 z-284)+138)-16 \big] f_0^{(3)}(z) \\
& \quad + (z-1) z^4 \big[ 3 \cc (z-1)+2 z (69 z-79)+38 \big] f_0^{(4)}(z) \\
& \quad + 3 (z-1)^2 (7 z-4) z^5 f_0^{(5)}(z)
        + (z-1)^3 z^6 f_0^{(6)}(z) = 0.
\end{split}
\end{align}
\end{subequations}
As discussed in the main text, we need to first ``change basis'' from the functions $f_i(z)$ to the $G_i(z)$, and then make an ansatz as a sum of bosonic blocks in order to solve the Casimir equations.
We start by expanding the external fields in terms of their conformal descendants, and we obtain the same expansion as in the right-hand side of~\eqref{eq:expansionDDDD_confPrim}.
Even though the operators at points $z_3$ and $z_4$ are longs, there are no new terms is the expansion because we work in the frame \fr 1, where $\theta_3 = \theta_4 = 0$, and therefore we can only get contributions from the superconformal primary field $A$.
As a result, the mapping~\eqref{eq:ansatz_change_basis_SSLL} is still valid, and we find that the change of basis must be given by~\eqref{eq:change_basis_SSSS} together with
\begin{align}
\label{eq:change_basis_SSLL}
\begin{split}
 f_9(z) 
& = 
    \frac{G_0(z)}{48 z^4}
  - \frac{\left(z^2+6\right) G_0'(z)}{288 z^3}
  - \frac{\left(5 z^2-12\right) G_0''(z)}{1152 z^2}
  + \frac{(z+4) (z-1) G_0^{(3)}(z)}{1152 z} \\
& + \frac{(z-1)^2 G_0^{(4)}(z)}{2304} 
  - \frac{G_1(z)}{24 z^4}
  - \frac{G_1'(z)}{144 z^2}
  - \frac{(z-1) G_1''(z)}{144 z^2}
  + \frac{G_2(z)}{6 z^4}.
\end{split}
\end{align}
The final step is to insert the change of basis~\eqref{eq:change_basis_SSSS} and~\eqref{eq:change_basis_SSLL} in the Casimir equations~\eqref{eq:casimirEq_SSSS} and~\eqref{eq:casimirEq_SSLL}, and use the resulting equations to fix the coefficients that apear in the ansatz~\eqref{eq:ansatz-sum-bosonic-blocks}. 
If we consider the block for an exchanged operator with quantum numbers $[\Delta, 0, 0]$, the solution to the equations is
\begin{align}
\label{eq:blocks_SSLL_00}
\begin{split}
& a_1 = \tfrac{1}{2} a_0 \left(\Delta-2\right), \\
& a_2 = -\tfrac{1}{16} a_0 \left(\Delta-3\right) \left(\Delta-2\right),
\end{split}
\begin{split}
& e_1 = -\tfrac{1}{2} e_0 \left(\Delta+3\right), \\
& e_2 = -\tfrac{1}{16} e_0 \left(\Delta+3\right) \left(\Delta+4\right),
\end{split}
\end{align}
and $b_i = c_i = d_i = 0$. 
Note that one of the free parameters, say $a_0$, can be fixed by choosing an overall normalization of the conformal block, as we did in~\eqref{eq:sol_SSSS_00}. 
However, the new feature is that there is still a free parameter $e_0$ that cannot be fixed by superconformal symmetry.

As a consistency check, we can take the OPE coefficients and norms of section~\ref{sec:blocks-two-three-pt-funs} to rederive this result.
The superblocks are given by
\begin{align}
\begin{split}
 & G_0(z) 
 = \frac{\lambda_{AAA} \tilde \lambda_{AAA}}{\langle A | A \rangle} \, g_{\Delta    }^{0,0}(z)
 + \frac{\lambda_{AAG} \tilde \lambda_{AAG}}{\langle G | G \rangle} \, g_{\Delta + 2}^{0,0}(z), \\
 & G_1(z) 
 = \frac{\lambda_{BBA} \tilde \lambda_{AAA}}{\langle A | A \rangle} \, g_{\Delta    }^{0,0}(z)
 + \frac{\lambda_{BBG} \tilde \lambda_{AAG}}{\langle G | G \rangle} \, g_{\Delta + 2}^{0,0}(z), \\
 & G_2(z) 
 = \frac{\lambda_{CCA} \tilde \lambda_{AAA}}{\langle A | A \rangle} \, g_{\Delta    }^{0,0}(z)
 + \frac{\lambda_{CCG} \tilde \lambda_{AAG}}{\langle G | G \rangle} \, g_{\Delta + 2}^{0,0}(z), \\
\end{split}
\end{align}
As in section~\ref{sec:blocks-two-three-pt-funs}, $\lambda_{O_1 O_2 O_3}$ denotes the OPE coefficient of two fields from the displacement multiplet with one operator from a long scalar multiplet, namely  $O_1, O_2 \in \DD$ and $O_3 \in \OO'$.
However, now one needs to consider also $\tilde \lambda_{O_1 O_2 O_3}$, where $O_1$ and $O_2$ are descendents of the external long $\OO$, but $O_3$ is a descendant of the exchanged long $\OO'$.
To recover equation~\eqref{eq:blocks_SSLL_00} we fix
\begin{align}
 a_0 = \frac{\lambda_{AAA} \tilde \lambda_{AAA}}{\langle A | A \rangle}, \qquad
 e_0 = \frac{\lambda_{AAG} \tilde \lambda_{AAG}}{\langle G | G \rangle}.
\end{align}
Then, for example, $a_1 = a_0 \lambda_{BBA} / \lambda_{AAA}$, and using~\eqref{eq:DDL_opecoefs} we recover the blocks~\eqref{eq:blocks_SSLL_00}. This works in an identical way for the other $a_i$ and $e_i$.

As in the case of four displacements, we can also have an exchange $[\Delta, 1, 0]$, with solution given by
\begin{align}
 c_1 = -\tfrac{1}{2} c_0, \qquad
 c_2 = \tfrac{1}{48} c_0 \left(\Delta - 2\right) \left(\Delta + 3\right),
\end{align}
where $a_i = b_i = d_i = e_i = 0$ and  we could fix the normalization of the block by $c_0 = 1$.

\subsection{Four longs}

Finally, we compute the superconformal blocks that appear in the four-point function of long scalar operators in the $(12)\to(34)$ channel
\begin{align}
\label{eq:expansion-superconf-blocks-LLLL}
 \langle \longM(z_1) \longM(z_2) \longM'(z_3) \longM'(z_4) \rangle 
 = \frac{1}{Z_{12}^{2\Delta_\OO} Z_{34}^{2\Delta_\OO'}} 
   \sum_{\OO''} \lambda_{\OO\OO\OO''} \lambda_{\OO'\OO'\OO''} \, \GG_{\OO''}(I_a),
\end{align}
where for simplicity we assume that $\Delta_1 = \Delta_2 = \Delta_\OO$ and $\Delta_3 = \Delta_4 = \Delta_\OO'$.
The steps in the calculation are very similar to the other studied cases, but the equations soon become quite long.
For this reason, we will skip some intermediate results in our presentation, but the interested reader can find the details in an attached \mathematica notebook.
The authors are also happy to provide further details on request.

First, we consider the four-point function of interest, which is given by~\eqref{eq:four-pt-fun-longs}, and act on it with the Casimir operator $\CC_{12}^2$.
Since we do not impose any shortening conditions to the four-point function, the full system of Casimir equations involves ten independent functions $f_0(z), \ldots, f_9(z)$.
The explicit differential equations, which are not particularly illuminating, can be found in the attached notebook.

In order to solve these equations, we need to first ``change basis'' to functions $G_i$ that capture the contribution of the conformal descendants in our multiplets.
In addition to~\eqref{eq:ansatz_change_basis_SSLL}, we need to make the following identifications:
\begin{align}
\begin{alignedat}{3}
& \langle E_{\ia\ib}^{\bA\bB}(x_1) E_{\ic\id}^{\bC\bD}(x_2) A(0) A(\infty) \rangle 
&& \to \;\;
&& \frac{\ee_{\ia\ib}\ee_{\ic\id} 
        (\ee^{\bA \bC} \ee^{\bB \bD} + \ee^{\bA \bD} \ee^{\bB \bC})
        }{|x_{12}|^{2\Delta_\OO + 2}} G_3(z), \\
& \langle A(x_1) \, G^{\text{p}}(x_2) \, A(0) \, A(\infty) \rangle 
&& \to \;\;
&& \frac{1}{|x_{12}|^{2\Delta_\OO+2}} \, G_4(z), \\
& \langle G^{\text{p}}(x_1) \, A(x_2) \, A(0) \, A(\infty) \rangle 
&& \to \;\;
&& \frac{1}{|x_{12}|^{2\Delta_\OO+2}} \, G_5(z), \\
& \langle B_\ia^\bA(x_1) \, F^{\text{p}}{}_\ib^\bB(x_2) \, A(0) \, A(\infty) \rangle 
&& \to \;\;
&& \frac{\ee^{\bA\bB} \ee_{\ia \ib}}{|x_{12}|^{2\Delta_\OO + 2}} \, G_6(z), \\
& \langle F^{\text{p}}{}^{\bA}_{\ia}(x_1) \, B_\ib^\bB(x_2) \, A(0) \, A(\infty) \rangle 
&& \to \;\;
&& \frac{\ee^{\bA\bB} \ee_{\ia \ib}}{|x_{12}|^{2\Delta_\OO + 2}} \, G_7(z), \\
& \langle F^{\text{p}}{}_{\ia}^\bA(x_1) \, F^{\text{p}}{}_{\ib}^\bB(x_2) \, A(0) \, A(\infty) \rangle 
&& \to \;\;
&& \frac{\ee^{\bA\bB} \ee_{\ia \ib}}{|x_{12}|^{2\Delta_\OO + 3}} \, G_8(z), \\
& \langle G^{\text{p}}(x_1) \, G^{\text{p}}(x_2) \, A(0) \, A(\infty) \rangle 
&& \to \;\;
&& \frac{1}{|x_{12}|^{2\Delta_\OO+4}} \, G_9(z).
\end{alignedat}
\end{align}
Here one needs to be careful to map the $G_i(z)$ with the true conformal descendants in the $\OO(z)$ superfield, namely one needs to use $F^{\text{p}}$ and $G^{\text{p}}$ defined in~\eqref{eq:true-conf-desc}.
With the above identifications, and following the obvious generalization of the steps in the main text, one can find the explicit change of basis $f_i(z) \to G_i(z)$.
Again, this transformation is a bit involved, and the interested reader can find it in the attached notebook.

Finally, we make an ansatz for the functions $G_i(z)$ as a finite sum of $\sl(2; \mathbb{R})$ blocks, as in equation~\eqref{eq:ansatz-sum-bosonic-blocks}.
Unlike the cases described so far, some of the $G_i$ represent four-point functions of descendants where the operators at $x_1$ and $x_2$ have different dimensions.
In these cases, the sum of bosonic blocks must be given by the blocks~\eqref{eq:chiral-block} with $\Delta_{12} \ne 0$.
More specifically, we use the ansatz
\begin{align}
\label{eq:ansatz-sum-bosonic-blocks-longs}
 G_i(z)
 = a_i \, g_{\Delta}^{\Delta_{12},0}(z)
 + b_i \, g_{\Delta+\tfrac{1}{2}}^{\Delta_{12},0}(z)
 + c_i \, g_{\Delta+1}^{\Delta_{12},0}(z)
 + d_i \, g_{\Delta+\tfrac{3}{2}}^{\Delta_{12},0}(z)
 + e_i \, g_{\Delta+2}^{\Delta_{12},0}(z),
\end{align}
where
\begin{align}
 \Delta_{12} = 
 \begin{cases} 
  -2 & \text{for } G_4(z) \\
  +2 & \text{for } G_5(z) \\
  -1 & \text{for } G_6(z) \\
  +1 & \text{for } G_7(z) \\
  0 & \text{otherwise}
 \end{cases}.
\end{align}

With these ingredients, one can fix the coefficients $a_i, \ldots, e_i$ by solving the Casimir equations as we previously did.
Before we present the solutions, let us make some comments.
First, compared to the cases studied before, there is a new solution corresponding to an exchanged operator with quantum numbers $[\Delta,0,1]$, namely without transverse spin but with \Rsymmetry.
Similarly to the discussion of the $\langle \DD \DD \OO \OO \rangle$ blocks, there are free parameters left in the solution.
Some of them can be fixed by choosing an appropriate normalization, but superconformal symmetry is not powerful enough to fix the rest.

\paragraph{Scalar exchange}

For an exchanged operator with quantum numbers $[\Delta, 0, 0]$,  the Casimir eigenvalue is $\cc = \Delta(\Delta+1)$ and the blocks are given by
\begin{align}
\begin{split}
 & a_1 = \frac{1}{2} a_0 \left(\Delta -2 \Delta_{\OO }\right), \\
 & a_3 = - \frac{1}{16} a_0 \left(\Delta -2 \Delta_{\OO }-1\right) \left(\Delta -2 \Delta_{\OO }\right) - a_2, \\
 & a_4 = + a_5 = \frac{a_0 \left(\Delta_{\OO }+2\right) \left(2 \Delta_{\OO }-\Delta \right) \left(-\Delta +2 \Delta_{\OO}+1\right)}{24 \left(2 \Delta_{\OO}+1\right)} + \frac{2 a_2}{3}, \\
 & a_6 = - a_7 = \frac{a_0 \left(\Delta_{\OO}+2\right) \left(2 \Delta_{\OO}-\Delta \right) \left(-\Delta +2 \Delta_{\OO}+1\right)}{6 \left(2 \Delta_{\OO}+1\right)} + \frac{8 a_2}{3}, \\
 & a_8=\frac{a_0 \left(2 \Delta_{\OO}-\Delta \right) \left(-\Delta +2 \Delta_{\OO}+1\right) \left(-\Delta +2 \Delta_{\OO}+2\right) \left(\Delta_{\OO}+2\right){}^2}{18 \left(2 \Delta_{\OO}+1\right){}^2} \\
 & \qquad +\frac{8 a_2 \left(-\Delta +2 \Delta_{\OO}+2\right)}{3 \left(2 \Delta_{\OO}+1\right)}, \\
 & a_9=\frac{a_0 \left(2 \Delta_{\OO}-\Delta \right) \left(-\Delta +2 \Delta_{\OO}+1\right) \left(-\Delta +2 \Delta_{\OO}+2\right) \left(-\Delta +2 \Delta_{\OO}+3\right) \left(\Delta_{\OO}+2\right){}^2}{576 \left(2 \Delta_{\OO}+1\right){}^2} \\
 & \qquad +\frac{a_2 \left(-\Delta +2 \Delta_{\OO}+2\right) \left(-\Delta +2 \Delta_{\OO}+3\right)}{12 \left(2 \Delta_{\OO}+1\right)},
\end{split}
\end{align}
and
\begin{align}
\begin{split}
 & e_1=\frac{1}{2} e_0 \left(-\Delta -2 \Delta_{\OO}-1\right), \\
 & e_3=-\frac{1}{16} e_0 \left(\Delta +2 \Delta_{\OO}+1\right) \left(\Delta +2 \Delta_{\OO}+2\right) - e_2, \\
 & e_4 = + e_5 = \frac{e_0 (\Delta +2) (\Delta +3) \left(\Delta_{\OO}+2\right) \left(\Delta +2 \Delta_{\OO}+1\right) \left(\Delta +2 \Delta_{\OO}+2\right)}{24 \Delta  (\Delta +1) \left(2 \Delta_{\OO}+1\right)} \\
 & \qquad \qquad \quad + \frac{2 (\Delta +2) (\Delta +3) e_2}{3 \Delta  (\Delta +1)}, \\
 & e_6 = -e_7 = -\frac{e_0 (\Delta +2) \left(\Delta_{\OO}+2\right) \left(\Delta +2 \Delta_{\OO}+1\right) \left(\Delta +2 \Delta_{\OO}+2\right)}{6 (\Delta +1) \left(2 \Delta_{\OO}+1\right)}-\frac{8 (\Delta +2) e_2}{3 (\Delta +1)}, \\
 & e_8=\frac{e_0 \left(\Delta +2 \Delta_{\OO}+1\right) \left(\Delta +2 \Delta_{\OO}+2\right) \left(\Delta +2 \Delta_{\OO}+3\right) \left(\Delta_{\OO}+2\right){}^2}{18 \left(2 \Delta_{\OO}+1\right){}^2} \\
 & \qquad +\frac{8 e_2 \left(\Delta +2 \Delta_{\OO}+3\right)}{3 \left(2 \Delta_{\OO}+1\right)}, \\
 & e_9=\frac{e_0 \left(\Delta +2 \Delta_{\OO}+1\right) \left(\Delta +2 \Delta_{\OO}+2\right) \left(\Delta +2 \Delta_{\OO}+3\right) \left(\Delta +2 \Delta_{\OO}+4\right) \left(\Delta_{\OO}+2\right){}^2}{576 \left(2 \Delta_{\OO}+1\right){}^2} \\
 & \qquad +\frac{e_2 \left(\Delta +2 \Delta_{\OO}+3\right) \left(\Delta +2 \Delta_{\OO}+4\right)}{12 \left(2 \Delta_{\OO}+1\right)},
\end{split}
\end{align}
with all other coefficients vanishing: $b_i = c_i = d_i = 0$.

\paragraph{Transverse-spin charged exchange}

For an exchanged operator with quantum numbers $[\Delta, 1, 0]$, the Casimir eigenvalue is $\cc = \Delta(\Delta + 1) + 2$ and the blocks are given by
\begin{align}
\begin{split}
 & c_1 = \frac{1}{2} c_0 \left(1-2 \Delta_{\OO}\right), \\
 & c_2 = \frac{1}{48} c_0 \big(\Delta(\Delta + 1) -6 \Delta_{\OO}(\Delta_{\OO}+1)+6\big), \\
 & c_3 = -\frac{1}{8} c_0 \left(\Delta_{\OO}-1\right){}^2, \\
 & c_4 = + c_5 = -\frac{c_0 (\Delta +1) (\Delta +2) \left(\Delta_{\OO}-1\right)}{24 \left(2 \Delta_{\OO}+1\right)}, \\
 & c_6 = - c_7 = -\frac{c_0 (\Delta +1) \left(\Delta_{\OO}-1\right)}{3 \left(2 \Delta_{\OO}+1\right)}, \\
 & c_8 = \frac{c_0 \left(\Delta_{\OO}-1\right){}^2 \left(2 \Delta_{\OO}+3\right) \left(2 \Delta_{\OO}-\Delta +1\right) \left(2 \Delta_{\OO}+\Delta +2\right)}{18 \left(2 \Delta_{\OO}+1\right){}^2}, \\
 & c_9 = \frac{c_0 \left(\Delta_{\OO}-1\right){}^2 \left(2 \Delta_{\OO}-\Delta +1\right) \left(2 \Delta_{\OO}-\Delta +2\right) \left(2 \Delta_{\OO}+\Delta +2\right) \left(2 \Delta_{\OO}+\Delta +3\right)}{576 \left(2 \Delta_{\OO}+1\right){}^2},
\end{split}
\end{align}
with all other coefficients vanishing: $a_i = b_i = d_i = e_i = 0$.

\paragraph{\Rsymmetry charged exchange}

Finally, when the exchanged operator has quantum numbers $[\Delta, 0, 1]$, the Casimir eigenvalue is $\cc = \Delta(\Delta + 1) - 4$ and the solution is
\begin{align}
\begin{split}
 & c_1 = -c_0 \left(\Delta_{\OO}+1\right), \\
 & c_2 = -\frac{1}{8} c_0 \left(\Delta_{\OO}+2\right){}^2, \\
 & c_3 = \frac{1}{48} c_0 \big(\Delta(\Delta+1) - 6 \Delta_{\OO}(\Delta_{\OO}+1) \big), \\
 & c_4 = c_5 = \frac{c_0 (\Delta +1) (\Delta +2) \left(\Delta_{\OO}+2\right)}{24 \left(2 \Delta_{\OO}+1\right)}, \\
 & c_6 = - c_7 = -\frac{c_0 (\Delta +1) \left(\Delta_{\OO}+2\right)}{6 \left(2 \Delta_{\OO}+1\right)}, \\
 & c_8 = \frac{c_0 \Delta_{\OO} \left(\Delta_{\OO}+2\right){}^2 \left(2 \Delta_{\OO}-\Delta +1\right) \left(2 \Delta_{\OO}+\Delta +2\right)}{9 \left(2 \Delta_{\OO}+1\right){}^2}, \\
 & c_9 = \frac{c_0 \left(\Delta_{\OO}+2\right){}^2 \left(2 \Delta_{\OO}-\Delta +1\right) \left(2 \Delta_{\OO}-\Delta +2\right) \left(2 \Delta_{\OO}+\Delta +2\right) \left(2 \Delta_{\OO}+\Delta +3\right)}{576 \left(2 \Delta_{\OO}+1\right){}^2},
\end{split}
\end{align}
with $a_i = b_i = d_i = e_i = 0$.

% \bibliography{./auxi/biblio.bib}
% \bibliographystyle{./auxi/JHEP}

\providecommand{\href}[2]{#2}\begingroup\raggedright\endgroup

\end{document}